\documentclass[twocolumn,tighten]{aastex63}
\usepackage{amssymb, amsmath,framed}

\usepackage{bm}
\expandafter\ifx\csname package@font\endcsname\relax\else
 \expandafter\expandafter
 \expandafter\usepackage
 \expandafter\expandafter
 \expandafter{\csname package@font\endcsname}
\fi
\def\bq{\begin{equation}}
\def\eq{\end{equation}}
\def\bqy{\begin{eqnarray}}
\def\eqy{\end{eqnarray}}

\def\p{\partial}

\def\p{\partial}

\def\R{\mathbb{R}}

 \def\q#1#2{q^{\hspace{1pt}#1}_{\,,\hspace{1pt}#2}}
 \def\a#1#2{a^{\hspace{1pt}#1}_{\,,\hspace{1pt}#2}}
 \def\d#1#2{\delta^{\hspace{1pt}#1}_{\,,\hspace{1pt}#2}}

\begin{document}
\title{\large{Propulsion of Spacecrafts to Relativistic Speeds Using Natural Astrophysical Sources}}

\correspondingauthor{Manasvi Lingam}
\email{mlingam@fit.edu}

\author{Manasvi Lingam}
\affiliation{Department of Aerospace, Physics and Space Sciences, Florida Institute of Technology, Melbourne FL 32901, USA}
\affiliation{Institute for Theory and Computation, Harvard University, Cambridge MA 02138, USA}

\author{Abraham Loeb}
\affiliation{Institute for Theory and Computation, Harvard University, Cambridge MA 02138, USA}

\begin{abstract}
In this paper, we explore from a conceptual standpoint the possibility of using natural astrophysical sources to accelerate spacecrafts to relativistic speeds. We focus on light sails and electric sails, which are reliant on momentum transfer from photons and protons, respectively, because these two classes of spacecrafts are not required to carry fuel on board. The payload is assumed to be stationed near the astrophysical source, and the sail is subsequently unfolded and activated when the source is functional. By considering a number of astrophysical objects such as massive stars, microquasars, supernovae, pulsar wind nebulae, and active galactic nuclei, we show that terminal speeds approaching the speed of light might be realizable under idealized circumstances provided that sufficiently advanced sail materials and control techniques exist. We also investigate the constraints arising from the sail's material properties, the voyage through the ambient source environment, and the passage through the interstellar medium. While all of these considerations pose significant challenges to spacecrafts, our analysis indicates that they are not insurmountable in optimal conditions. Finally, we sketch the implications for carrying out future technosignature searches.\\
\end{abstract}

\section{Introduction} \label{SecIntro}
The 1950s-1970s witnessed an unprecedented investment of time, money and resources in developing space exploration as part of the space race \citep{McDo85,Bur98,Neu18}, but the decades that followed proved to be more fallow \citep{NSM08,McCu11,Bri19}. In recent times, however, there has been a renewed interest in the resumption of space exploration. NASA has announced their intentions to return humans to the Moon,\footnote{\url{https://www.nasa.gov/specials/apollo50th/back.html}} and thereafter land people on Mars in the near future \citep{Sz19}.\footnote{\url{https://www.nasa.gov/sites/default/files/atoms/files/nationalspaceexplorationcampaign.pdf}} In parallel, a number of private companies such as Space X have also proclaimed their plans to make humanity a ``multi-planetary species'' \citep{Z11,Gen14,Musk17,Dav18,DAP20}. 

In light of the renewed interest in space exploration, increasing attention is being devoted to modeling new propulsion systems \citep{Fris03,Long11}. Whilst chemical rockets still remain the \emph{de facto} mode of space exploration, they are beset by a number of difficulties. First and foremost, their necessity of having to transport fuel on board imposes prohibitive requirements on their mass and economic cost. Second, by virtue of the rocket equation, they are severely hampered in terms of the maximum speeds that they can reach. As a result, numerous alternative technologies are being seriously pursued that do not require the on-board transport of fuel \citep{Taj12}. Examples in this category include light sails \citep{Zand24,Forw84,McIn04,Vul12,Lub16,FSE16}, magnetic sails \citep{ZA91,Djo18} and electric sails \citep{Jan04,JTP10}.

When it comes to interplanetary travel within the inner Solar system, speeds of order of tens of km/s suffice to undertake space exploration over a human lifetime. However, in the case of interstellar travel, there are significant benefits that arise from developing propulsion technologies that are capable of attaining a fraction of the speed of light. The recently launched \emph{Breakthrough Starhot} initiative is a natural example, because it aims to send a gram-sized spacecraft to Proxima Centauri at $20\%$ the speed of light by employing a laser-driven light sail \citep{Pop17,WDK18}.\footnote{\url{https://breakthroughinitiatives.org/initiative/3}} Setting aside the technical challenges, one of the striking aspects of this mission is the energetic cost that it entails: the laser array that accelerates the light sail must have a peak transmission power of $\sim 100$ GW \citep{Park18}. 

Hence, this immediately raises the question of whether it is feasible to harness \emph{natural} astrophysical sources to achieve relativistic speeds to undertake interstellar travel \citep{McC93,Loeb}. The technological viability and the accompanying pros and cons of interstellar travel have been extensively debated \citep{FJ85}; a summary of the benefits arising from interstellar travel can be found in \citet{Craw14,Zub19}. Fortunately, the universe is replete with high-energy astrophysical phenomena. Many of them are known to be highly efficient at accelerating particles such as electrons, protons and even dust to relativistic speeds \citep{RB07,Mel09,MSL11,Draine,HLS15}. Likewise, it ought to be feasible to tap these sources and drive spacecrafts to relativistic speeds. Not only does it have the advantage of potentially cutting costs for technological species but it may also lower their likelihood of being detected because propulsion via laser arrays engenders distinctive technosignatures \citep{GL15,BB16,LL17}.

In this paper, we investigate whether it is feasible to utilize natural astrophysical sources to achieve high terminal speeds, which can approach the speed of light in some cases. We will study two different classes of propulsion systems herein: light sails in Sec. \ref{SecLS}, and electric sails in Sec. \ref{SecESail}. For both propulsion systems, we suppose that the payload is parked at the initial distance from the source with its sail folded and the latter is unfurled at the time of launch (i.e., when the object becomes active). In other words, the spacecraft must travel from its parent system to a suitable high-energy astrophysical object and position itself there. 

While the journey to the source may take a long time (e.g., on the order of $10^5$ yrs), the ensuing advantage is that subsequent interstellar travel would be relativistic and does not entail further energy expenditure, because the acceleration is provided for ``free'' by the source. Once the acceleration phase is over, the sail would be folded back to reduce damage and friction, with the payload designed such that its cross-sectional area parallel to the direction of the motion is minimized. We conclude with a summary of our central results and the limitations of our analysis, and we briefly delineate the ramifications for detecting technosignatures in Sec. \ref{SecConc}.

\section{Light Sails}\label{SecLS}
We will investigate the prospects for accelerating light sails to high speeds using astrophysical sources.

\subsection{Terminal velocity of relativistic light sails}\label{SSecRelVTLS}
Although we will deal with weakly relativistic light sails for the majority of our analysis, it is instructive to tackle the relativistic case first; this scenario was first modeled by \citet{Marx66}. For a light sail powered by an isotropic astrophysical source of constant luminosity ($L_\star$),\footnote{This approximation represents an idealized limit because all astrophysical sources are characterized by temporal variability. For instance, flares and superflares on active stars are responsible for transiently boosting the luminosity \citep{LiLo,YMA19,Lin19,GZS}.} and supposing that the sail reflectance is close to unity ($R \approx 1$), the corresponding equation of motion is derivable from Equation (2) of \citet{MVP09} and Equation (9) of \citet{KLZ18}:
\begin{equation}\label{RelEOM}
    \gamma^3 \frac{d \beta}{dt} \approx \frac{L_\star}{2\pi r^2 \Sigma_s c^2} \left(\frac{1-\beta}{1+\beta}\right),
\end{equation}
where $\beta = v/c$, $\gamma = 1/\sqrt{1-\beta^2}$, and $\Sigma_s$ is the mass per unit area of the sail; we adopt the fiducial value of $\Sigma_0 \approx 2 \times 10^{-4}$ kg/m$^2$ as it could be feasible in the near-future \citep{Park18}. Note that $v$ denotes the instantaneous velocity of the sail, and $r$ represents the time-varying distance between the sail and the astrophysical source.

In formulating this equation, we have not accounted for the inward gravitational acceleration, but this term is negligible provided that $L_\star \gtrsim 0.01 L_\odot$ \citep{LL20}. Likewise, the drag force has been neglected, as it does not alter the results significantly in the limits of $\beta \rightarrow 0$ and $\beta \rightarrow 1$ \citep{Ho17}. We have also presumed that the light sail preserves a constant orientation relative to the source at all points during its acceleration. This requires the selection of suitable sail architecture \citep{ML17} as well as the deployment of nanophotonic structures for self-stabilization \citep{IA19}. Lastly, we implicitly work with the scenario in which the payload mass ($M_{pl}$) is distinctly smaller than, or comparable to at most, to the sail mass ($M_s$).

Next, after employing the relation $dt = dr/(\beta c)$ and integrating (\ref{RelEOM}), we arrive at
\begin{equation}\label{RelV}
    \frac{1}{3} \left[1 + \frac{\sqrt{1+\beta_T}\left(-1+2\beta_T\right)}{\left(1-\beta_T\right)^{3/2}}\right] \approx  \frac{L_\star}{2\pi \Sigma_s c^3 d_0},
\end{equation}
where $d_0$ represents the initial distance from the source (i.e., when the light sail is launched) and $\beta_T$ is the normalized terminal velocity achieved by the light sail. Instead of calculating $\beta_T$, it is more instructive to express our results in terms of the spatial component of the 4-velocity, namely, $U_T = \beta_T \gamma_T$ because $U_T \rightarrow \beta_T$ for $\beta_T \ll 1$ and $U_T \rightarrow \gamma_T$ for $\beta_T \rightarrow 1$. 

The next aspect to consider is the initial launch distance. While this appears to be a free parameter, it will be constrained by thermal properties in reality \citep[Chapter 2.6]{McIn04}. We introduce the notation $\varepsilon$ for the sbsorptance (note that $\varepsilon \ll 1$ under optimal circumstances) and denote the sail temperature at the initial location by $T_s$. If we suppose that the sail behaves as a blackbody, we obtain
\begin{equation}\label{HeatRel}
    \frac{\varepsilon L_\star}{4 \pi d_0^2} \approx \sigma T_s^4, 
\end{equation}
provided that the total sail emissivity associated with its front and back sides is close to unity.\footnote{If the emissivity is distinctly lower than unity, our calculations still remain valid after implementing the transformation $\varepsilon \rightarrow \varepsilon/\psi$, where $\psi$ is the total emissivity of the sail.} As we shall hold $T_s$ fixed henceforth at $d_0$, it is reasonable to presume that the initial emissivity - which is dependent on $T_s$ \citep[Equation 9]{AK17} - is crudely invariant for all astrophysical sources. This equation can be duly inverted to solve for $d_0$, thus yielding
\begin{equation}\label{dzero}
    d_0 \approx 0.17\,\mathrm{AU}\,\left(\frac{\varepsilon}{0.01}\right)^{1/2} \left(\frac{L_\star}{L_\odot}\right)^{1/2} \left(\frac{T_s}{300\,\mathrm{K}}\right)^{-2}.
\end{equation}
We have introduced the fiducial values of $\varepsilon \approx 0.01$ and $T_s \approx 300$ K. The normalization factor for $\varepsilon$ is optimistic because this value embodies the aggregate across all wavelengths, but it might be realizable through the use of multilayer stacking techniques \citep[Figure 3]{ADI18}. The temperature of $300$ K was chosen on the premise that it represents a comfortable value for organic lifeforms as well as electronic instrumentation. After combining (\ref{HeatRel}) and (\ref{RelV}), the latter is expressible as
\begin{equation}\label{Relbeta}
    \frac{1}{3} \left[1 + \frac{\sqrt{1+\beta_T}\left(-1+2\beta_T\right)}{\left(1-\beta_T\right)^{3/2}}\right] \approx  \frac{T_s^2}{\Sigma_s c^3} \sqrt{\frac{L_\star \sigma}{\pi \varepsilon}},
\end{equation}
and upon substituting the previously specified parameters, the above equation simplifies to
\begin{eqnarray}\label{RelVTerm}
 && \frac{L_\star}{L_\odot} \approx 5.8 \times 10^{11}\,\left(\frac{\varepsilon}{0.01}\right)\left(\frac{\Sigma_s}{\Sigma_0}\right)^{2}\left(\frac{T_s}{300\,\mathrm{K}}\right)^{-4} \nonumber \\
 &&\hspace{0.5in} \times \,\left[1 + \frac{\sqrt{1+\beta_T}\left(-1+2\beta_T\right)}{\left(1-\beta_T\right)^{3/2}}\right]^2.
\end{eqnarray}

If we know the terminal speed that we wish to achieve using a suitable astrophysical source, we can employ this equation to estimate the requisite luminosity of the object. Before proceeding further, it is useful to consider two limiting cases. First, in the non-relativistic regime corresponding to $\beta_T \ll 1$, we obtain
\begin{equation}\label{NonrelTV}
  \frac{L_\star}{L_\odot} \approx 1.3 \times 10^{12}\,\beta_T^4 \left(\frac{\varepsilon}{0.01}\right)\left(\frac{\Sigma_s}{\Sigma_0}\right)^{2}\left(\frac{T_s}{300\,\mathrm{K}}\right)^{-4}.
\end{equation}
Next, if we consider the ultrarelativistic regime wherein $\gamma_T \gg 1$, we find that (\ref{RelVTerm}) reduces to
\begin{equation}
  \frac{L_\star}{L_\odot} \approx 9.3 \times 10^{12}\,\gamma_T^6 \left(\frac{\varepsilon}{0.01}\right)\left(\frac{\Sigma_s}{\Sigma_0}\right)^{2}\left(\frac{T_s}{300\,\mathrm{K}}\right)^{-4}.
\end{equation}
Hence, anticipating later results, it is evident that attaining the ultrarelativistic regime is very difficult because it necessitates $L_\star \gg 10^{13}\,L_\odot$. 

\begin{figure}
\includegraphics[width=7.5cm]{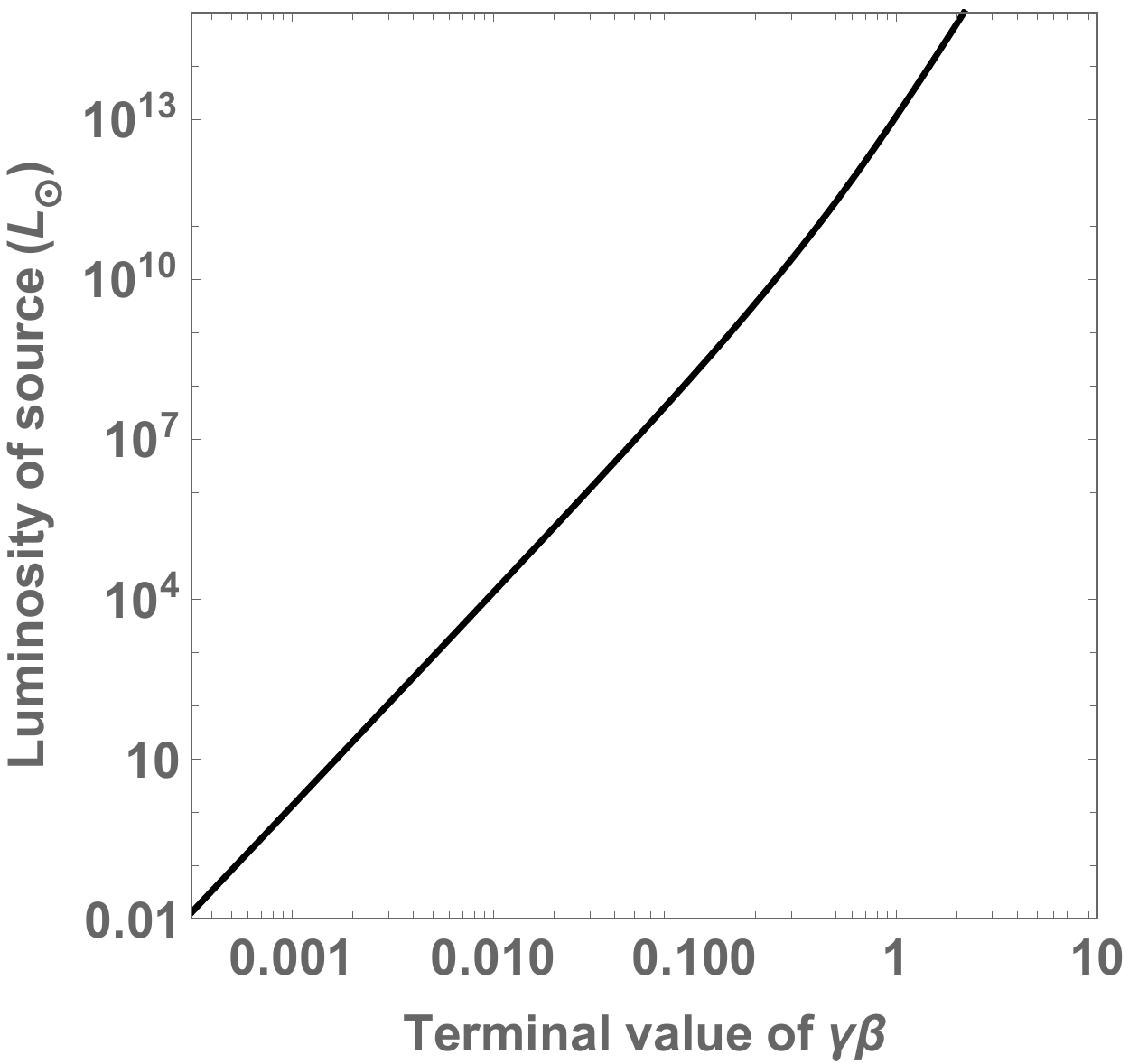} \\
\caption{The luminosity of the source (units of $L_\odot$) as a function of the terminal value of $\gamma \beta$, with the other free parameters in (\ref{RelVTerm}) held fixed at their fiducial values.}
\label{FigVelTermRel}
\end{figure}

In Fig. \ref{FigVelTermRel}, we have plotted the luminosity of the astrophysical source as a function of $U_T$. We have restricted the lower bound to $0.01 L_\odot$ because gravitational acceleration becomes important below this luminosity as noted previously, and the upper bound has been chosen based on the most luminous quasars. In the case of $U_T \ll 1$, we observe that the luminosity requirements are relatively modest. For example, we find that $L_\star \approx L_\odot$ leads to $\beta_T \approx 10^{-3}$. However, once we approach the regime of $U_T \sim 1$, the associated luminosity becomes very large, eventually exceeding that of virtually all known astrophysical objects. By inspecting the figure, it is observed that the plot behaves as a power law with an exponent of $+4$ up to $U_T \gtrsim 0.1$, as expected from (\ref{NonrelTV}).

\subsection{Terminal speeds of light sails powered by astrophysical sources}\label{SSecLSTermV}

At this point, it is useful to address some long-lived astrophysical sources in more detail. First, we consider the hottest and most massive stars in the Universe, whose luminosity can be roughly approximated by the Eddington luminosity \citep[Equation 22.10]{KWW12}. When expressed in terms of the stellar mass $M_\star$, the luminosity is given by
\begin{equation}\label{EddLum}
    L_\star \approx 3.8 \times 10^4\,L_\odot \left(\frac{M_\star}{M_\odot}\right).
\end{equation}
Hence, upon specifying $M_\star \approx 200\, M_\odot$, given that it seems characteristic of certain massive Wolf-Rayet stars in the Large Magellanic Cloud, the above scaling yields $L_\star \approx 7.6 \times 10^6\,L_\odot$ and thereby evinces reasonable agreement with observations \citep{HRT14,CCB16}. From Fig. \ref{FigVelTermRel}, we find that this luminosity yields a terminal speed of $\beta_T \approx 0.05$. 

Along similar lines, final speeds of $\sim 0.01 c$ are attainable by light sails near low-mass X-ray binaries because these objects have bolometric luminosities of $\lesssim 10^6 L_\odot$; these objects have the additional benefit of being long-lived, as their lifespans can reach $\sim 0.1$ Gyr \citep{Gil04}. Another class of objects that give rise to similar speeds are a particular category of X-ray binaries, known colloquially as microquasars \citep{Beck08}. As these sources comprise black holes with masses of $\sim 1$-$10 M_\odot$ \citep{Mir01,CSF05}, the use of (\ref{EddLum}) suggests that their typical luminosities are on the order of $10^5-10^6 L_\odot$, thereby giving rise to $\beta_T \sim 0.01$.

The next class of objects to consider are Active Galactic Nuclei (AGNs), whose luminosities are estimated via (\ref{EddLum}); the only difference is that $M_\star$ should be replaced with the mass ($M_\mathrm{BH}$) of the supermassive black hole \citep{Kro99}. As per theory and observations,  $M_\mathrm{BH} \sim 10^{11}\,M_\odot$ constitutes an upper bound for supermassive black holes in the current universe \citep{MMG11,IH16,PNF17,DGK17,IVH}. When this limit is substituted into (\ref{EddLum}) after invoking the fact that the Eddington factor is typically around unity during the quasar phase \citep{MRG04},\footnote{An Eddington ratio of unity may represent an optimistic choice \citep{KWF10,VB12}, but it permits us to gauge the maximum value of $\beta_T$ for a given AGN luminosity.} we find $L_\star \sim 3.8 \times 10^{15}\,L_\odot$. By plugging this value into (\ref{RelVTerm}), we end up with $\gamma_T \approx 2.9$. In other words, the most luminous AGNs are capable of driving light sails into the relativistic regime, but not to ultrarelativistic speeds.

Next, we turn our attention to supernovae (SNe). There are many categories of supernovae, each powered by different physical mechanisms, owing to which the identification of a characteristic luminosity is rendered difficult. A general rule of thumb is to assume a peak luminosity of $10^9 L_\odot$ \citep[Chapter 1]{BW17}, which yields $\beta_T \approx 0.15$ after making use of (\ref{RelVTerm}); in other words, typical SNe may accelerate light sails to mildly relativistic speeds \citep{Loeb}. It is, however, important to recognize that a special class of supernovae, known as superluminous supernovae (SLSNe), have peak luminosities that are $\gtrsim 100$ times larger than normal events \citep{GY19,Ins19}. Calculations based on numerical simulations and empirical data suggest that the upper bound on the peak luminosity of SLNe is approximately $5.2 \times 10^{12}\,L_\odot$ \citep{SW16}. By applying (\ref{RelVTerm}), we obtain $\beta_T \approx 0.66$, thereby suggesting that extreme SLSNe could accelerate light sails to significantly relativistic speeds.

\begin{table}
\caption{Terminal momentum per unit mass achievable by light sails near astrophysical objects}
\label{TabLS}
\begin{tabular}{|c|c|}
\hline 
Source & Terminal momentum ($\gamma \beta$)\tabularnewline
\hline 
\hline 
Sun & $\sim 10^{-4}$\tabularnewline
\hline 
Massive stars & $\sim 0.01$-$0.1$\tabularnewline
\hline 
Low-mass X-ray binaries & $< 0.01$\tabularnewline
\hline 
Microquasars & $< 0.01$\tabularnewline
\hline 
Supernovae & $\lesssim 0.1$-$1$\tabularnewline
\hline 
Active Galactic Nuclei & $\lesssim 1$\tabularnewline
\hline 
Gamma-ray bursts & $< 10$\tabularnewline
\hline 
\end{tabular}
\medskip

{\bf Notes:} $\gamma \beta$ denotes the terminal momentum per unit mass. It is important to recognize that this table yields the \emph{maximum} terminal momentum per unit mass attainable by light sails. In actuality, however, some of the sources will either be too transient (e.g., GRBs) to achieve the requisite speeds or manifest high particle densities that may cause damage to light sails; these issues are further analyzed in Secs. \ref{SSecAccLS} and \ref{SSecSourceC}. Based on these reasons, the above terminal momenta should be regarded as optimistic upper bounds; for more details, consult Secs. \ref{SSecLSTermV} to \ref{SSecISM}.
\end{table}

\subsection{Acceleration time for weakly relativistic light sails}\label{SSecAccLS}
The previous consideration of SNe brings up a crucial caveat that merits further scrutiny. Hitherto, we have implicitly operated under two implicit assumptions concerning the astrophysical object: (i) it has a constant luminosity ($L_\star$), and (ii) it remains functional for a sufficiently long time to effectively enable the light sail to attain speeds that are close to the terminal value calculated in \ref{RelVTerm}). It is apparent that these two assumptions will be violated for objects that are highly luminous, but remain so only for a transient period of time; examples of such objects are SNe and gamma-ray bursts (GRBs). In contrast, massive stars and AGNs are functional over long timescales ($\gtrsim 10^6$ yr).

\begin{figure}
\includegraphics[width=7.5cm]{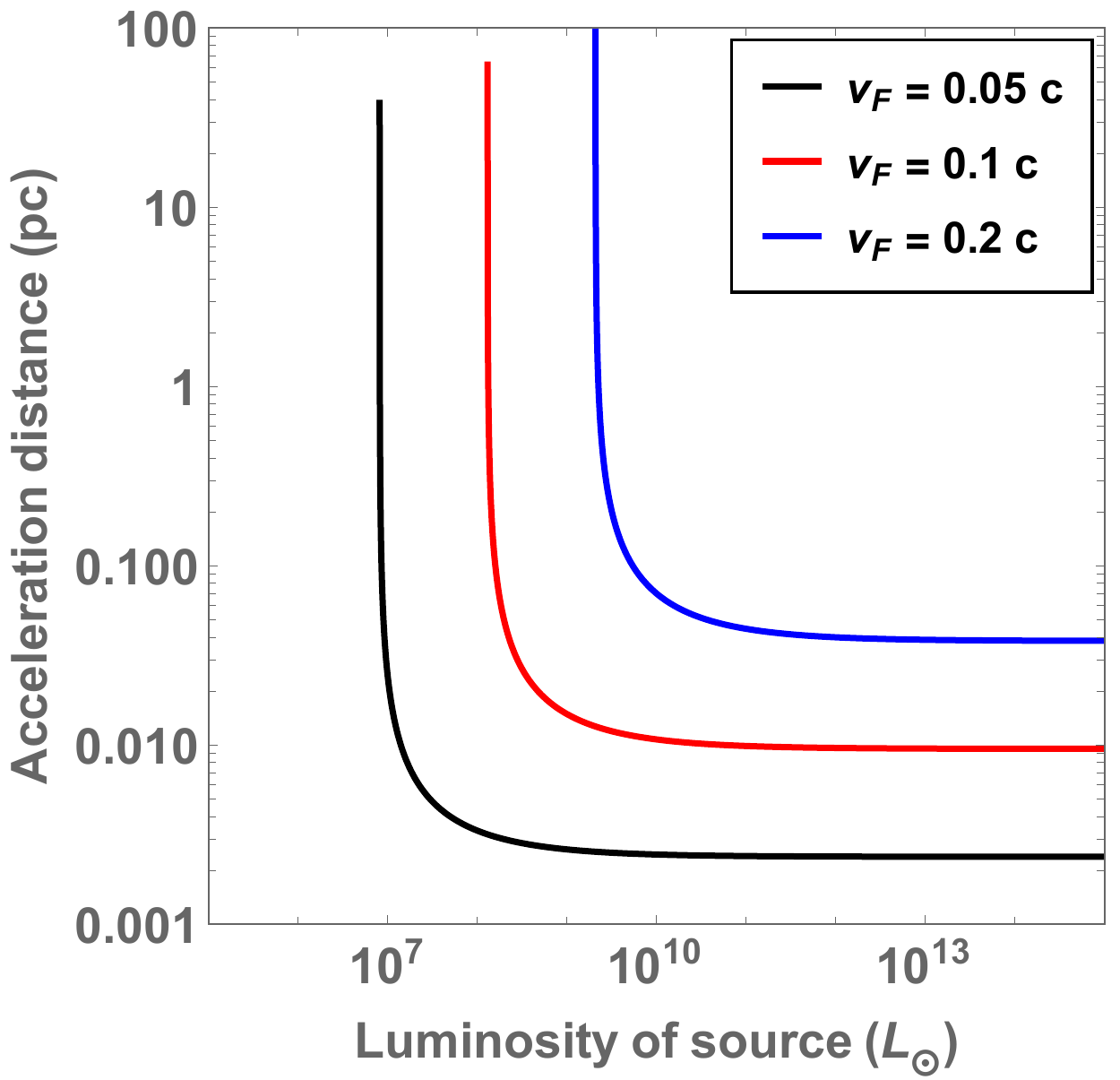} \\
\caption{Distance travelled by the light sail (units of pc) to achieve the desired final velocity ($v_F$) as a function of the luminosity of the source (units of $L_\odot$) using (\ref{AccD}). The red, black and blue curves correspond to different choices of $v_F$, while the other parameters are held fixed at their nominal values in (\ref{AccD}).}
\label{FigAccDist}
\end{figure}

Hence, it becomes necessary to address another major question: What is the time required for a light sail to achieve a desired final velocity ($v_F$)? We will adopt $v_F \sim 0.1 c$ because this is close to the terminal speeds associated with several high-energy astrophysical phenomena as well as comparable to the speed of laser-powered light sails such as \emph{Breakthrough Starshot}. Moreover, as this speed is weakly relativistic, it is ostensibly reasonable to utilize the non-relativistic counterpart of (\ref{RelEOM}) without the loss of much accuracy \citep[Chapter 7.3]{McIn04}. Upon integrating the non-relativistic version of (\ref{RelEOM}), by taking the limit $\beta \ll 1$, we get
\begin{equation}\label{NRVelEOM}
    v^2 \approx \frac{L_\star}{\pi c \Sigma_s d_0} \left(1 - \frac{d_0}{r}\right).
\end{equation}
Hence, the distance covered by the light sail before it attains the desired speed of $v_F$ is defined as $\Delta r = r_F-d_0$, where $r_F$ is the location at which $v = v_F$ is attained. Hence, upon further simplification, we end up with
\begin{equation}
    \Delta r \approx d_0 \left[\frac{L_\star}{\pi d_0 c^3 \beta_F^2 \Sigma_s} - 1\right]^{-1},
\end{equation}
where we have introduced the notation $\beta_F = v_F/c$. By making use of (\ref{HeatRel}), the above equation reduces to
\begin{eqnarray}\label{AccD}
&& \Delta r \approx 0.17\,\mathrm{AU}\,\left(\frac{\varepsilon}{0.01}\right)^{1/2} \left(\frac{L_\star}{L_\odot}\right)^{1/2} \left(\frac{T_s}{300\,\mathrm{K}}\right)^{-2} \nonumber \\
&& \hspace{0.2in} \times\,\Bigg[8.8 \times 10^{-5} \left(\frac{T_s}{300\,\mathrm{K}}\right)^{2} \left(\frac{\Sigma_s}{\Sigma_0}\right)^{-1} \left(\frac{\varepsilon}{0.01}\right)^{-1/2} \nonumber \\
&& \hspace{0.55in} \times \left(\frac{\beta_F}{0.1}\right)^{-2} \left(\frac{L_\star}{L_\odot}\right)^{1/2} - 1 \Bigg]^{-1}.
\end{eqnarray}
It is apparent from inspecting the above equation that $\Delta r > 0$ necessitates very high luminosities. This requirement is expected, because Fig. \ref{FigVelTermRel} illustrates that reaching a terminal speed on the order of $0.1 c$ is feasible only for highly luminous sources. We have plotted $\Delta r$ as a function of $L_\star$ in Fig. \ref{FigAccDist}. To begin with, we notice that $\Delta r > 0$ only for sufficiently high luminosities as explained previously. Second, at large luminosities, it is found that $\Delta r$ becomes independent of $L_\star$. This trend is discernible from (\ref{AccD}) after assuming that the first term inside the square brackets is much larger than unity.

It is convenient to define the following variable for the ensuing analysis:
\begin{eqnarray}\label{vinf}
&& v_\infty \equiv \sqrt{\frac{L_\star}{\pi c \Sigma_s d_0}} \approx 9.4 \times 10^{-4}\,c\,\left(\frac{L_\star}{L_\odot}\right)^{1/4} \left(\frac{\varepsilon}{0.01}\right)^{-1/4} \nonumber \\
&& \hspace{1.2in} \times\,\left(\frac{T_s}{300\,\mathrm{K}}\right)\left(\frac{\Sigma_s}{\Sigma_0}\right)^{-1/2}
\end{eqnarray}
By integrating (\ref{NRVelEOM}) and invoking the definition of $v_\infty$, we end up with
\begin{equation}\label{NRelREOM}
  r \sqrt{1 - \frac{d_0}{r}} + d_0\,\mathrm{tanh}^{-1}\left(\sqrt{1 - \frac{d_0}{r}}\right) \approx  v_\infty t.
\end{equation}
In particular, we are interested in calculating $\Delta t$, which is defined as the time at which $r = r_F$ and $v = v_F$. This timescale is determined by substituting $r = r_F$ in (\ref{NRelREOM}), but the final expression proves to be tedious (albeit straightforward to calculate), owing to which the explicit formula is not provided herein. 

\begin{figure}
\includegraphics[width=7.5cm]{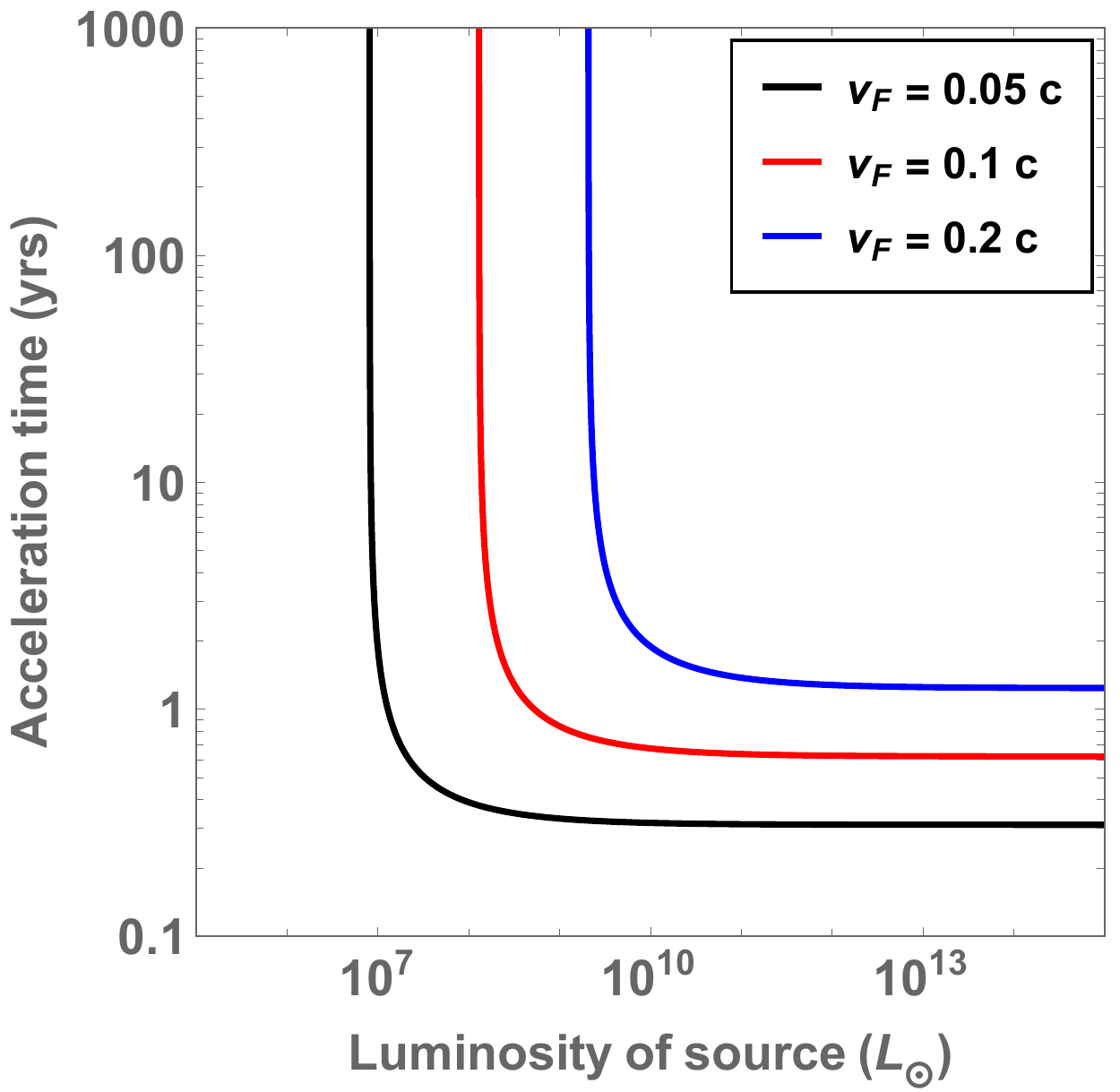} \\
\caption{Time required by the light sail (units of years) to achieve the desired final velocity ($v_F$) as a function of the luminosity of the source (units of $L_\odot$) using (\ref{NRelREOM}). The red, black and blue curves correspond to different choices of $v_F$, while the other parameters are held fixed.}
\label{FigAccTime}
\end{figure}

Fig. \ref{FigAccTime} shows $\Delta t$ as a function of $L_\star$ for different choices of $v_F$. We observe that $\Delta t$ is initially large but it rapidly reaches an asymptotic value, which is independent of $L_\star$. By considering the formal mathematical limit of $L_\star \rightarrow \infty$ and employing standard asymptotic techniques \citep{Olv74}, one arrives at $\Delta t \sim 2 d_0 v_F/v_\infty^2$. After using (\ref{dzero}) and (\ref{vinf}) in this asymptotic expression for $\Delta t$, we find that the dependence on $L_\star$ cancels out, thereby providing the explanation as to why $\Delta t$ attains a value independent of $L_\star$ in Fig. \ref{FigAccTime}. 

From an inspection of Fig. \ref{FigAccTime}, it is evident that AGNs comfortably satisfy the requirements for $\Delta t$ because they are typically active over timescales comparable to the Salpeter time, which has characteristic values of $\sim 10$-$100$ Myr \citep{Shen13}. In the case of SNe, we see that $\Delta t \sim 0.6$ yr is necessary to achieve a speed of $\sim 0.1$ c, but this number can be lowered further by tuning the other parameters; for example, increasing the temperature by $30\%$ yields $\Delta t \sim 77$ days. This estimate is comparable to the typical peak luminosity timescale for most classes of SNe, which is potentially a few months \citep{SW16}. Hence, it is conceivable that the timescale over which SNe are operational suffices to power light sails to weakly relativistic speeds. 

The situation is rendered very different, however, when we consider GRBs. In theory, the peak luminosities of GRBs are sufficiently high to enable $U_T \gg 1$ to be achieved in accordance with (\ref{RelVTerm}) and Fig. \ref{FigVelTermRel}. This is because most GRBs that have been detected are characterized by peak values of $L_\star > 10^{16} L_\odot$, although low-luminosity GRBs have also been identified \citep{ZZS18}. However, the real bottleneck is the timescale over which these phenomena are active: even the ultra-long GRBs have timescales of $\sim 10^4$ s \citep{GSA13,KZ15}. Hence, upon comparison with Fig. \ref{FigAccTime}, we see that this timescale is insufficient to accelerate the light sails to $\sim 0.1 c$. In addition, the close proximity of light sails to GRBs may cause damage to instruments and putative biota due to the high fluxes of ionizing radiation \citep{MT11}. Finally, as explained in the subsequent sections, higher values of $\Sigma_s$ and $\varepsilon$ are potentially necessary at these wavelengths, thereby suppressing $\beta_T$ by orders of magnitude.

\subsection{Sail properties: astrophysical constraints}\label{SSecSailM}
Broadly speaking, our model is characterized by the existence of three control parameters. Of the trio, a conservative choice was adopted for $T_s$, the sail temperature at the launch location. In fact, choosing $T_s \approx 400$ K would enable the attainment of higher terminal speeds, and not cause much damage to silicon-based electronics in the process. The damage to organic lifeforms could be more pronounced, but several authors have suggested that technological entities capable of interstellar travel may be post-biological in nature \citep{Fri80,Dick,Dick08,Sma12}, in which case the significance of this limitation would be diminished.

The other two parameters are the area density ($\Sigma_s$) and absorptance ($\varepsilon$). In our calculations, we have normalized $\Sigma_s$ by $\Sigma_0$, which effectively amounts to a sail thickness of $\lesssim 0.1$ $\mu$m. The astrophysical sources we have considered herein, however, span a wide range of luminosities and exhibit correspondingly different spectral energy distributions (SEDs). Thus, the challenge is to design sails that have low absorptance (i.e., high reflectance) while also maintaining a sail thickness compatible with $\Sigma_0$. As the terminal speed decreases monotonically with $\Sigma_s$ and $\varepsilon$, we emphasize that the results derived earlier represent optimistic upper bounds that are probably not realizable in practice. Furthermore, the SEDs of most astrophysical objects are broadband, unlike artificial (i.e., laser or maser) sources, thus making sail specification and design a challenging endeavor. Hence, unless advanced technology can make light sails effective across the range where a substantial fraction of photon output occurs, they will be unable to reach the high speeds obtained in the previous sections.

Studies dealing with reflectance in the $\gamma$-ray regime are relatively few in number. Hence, it is instructive to focus on X-rays as they represent the adjacent range. In the case of Si and SiO$_2$, two materials considered for \emph{Breakthrough Starshot}, the reflectance ($R$) has ranged between $\sim 0.1$-$1$ at wavelengths of $\lesssim 100\,\mathrm{\AA}$ \citep{TLM}; the issue, however, is that the corresponding grazing angles were low ($0$-$10^\circ$), implying that the sail would have to maintain an unusual orientation continuously with respect to the incoming radiation. At low grazing angles and wavelengths of $\sim 1$ nm, peak reflectances of $R \approx 0.2$-$0.3$ have been obtained for various multilayers \citep{CNY,SS99,VGW15,BSD16}. 

Although most of the prior studies were carried out at low grazing angles, several experimental studies indicate that X-ray mirrors with $R \gtrsim 0.2$ are realizable at near-normal incidence, albeit at \emph{select} wavelengths of typically $\sim 1$-$10$ nm \citep{BH82,TBC87,SRV91,MKS96,MSG98,EJH03,LCJ07,MSG09,LNS18}; multilayers comprised the likes of W/C, Ni/Ti, W/B$_4$C, Cr/Sc, Si/Mo, and Mo/Y. In most of the cases referenced hitherto, the total thickness of the multilayer was comparable to the thickness associated with $\Sigma_0$.

Looking beyond soft X-rays, we note that high reflectances have been achieved even for hard X-rays and soft $\gamma$-rays (whose energies are $\gtrsim 10$ keV) under very specific circumstances. \citet{SSB11} demonstrated via experiments and theory that $\sim 1$ mm thick diamond crystals were capable of achieving $>99\%$ reflectivity for photons of energies $13.7$ and $23.9$ keV via Bragg diffraction and backscattering at normal incidence; in general, the typical thickness of the crystal ought to be $\gtrsim 50$ $\mu$m to permit near-complete reflection \citep{SL12}. Another option to achieve $R \rightarrow 100\%$ is to ensure that the diamond crystal is perpetually positioned close to grazing incidence \citep{SSC10}.

Hitherto, we have discussed only the reflectance, but the results for the absorptance are complementary; if the transmittance is minimal, we have $\varepsilon = 1 - R$. In other words, when $R$ is much smaller than unity, we note that $\varepsilon \rightarrow 1$. Although this translates to an increase in $\varepsilon$ by two orders of magnitude compared to the fiducial choice, this does not pose a major thermal issue. The reason stems from (\ref{HeatRel}) and (\ref{dzero}), which imply that the launch distance is chosen such that the initial sail temperature $T_s$ is fixed at $\sim 300$ K. Hence, if $\varepsilon$ is elevated for certain astrophysical SEDs and sail materials, we find that $d_0$ is correspondingly increased, and vice-versa. In all cases, however, the sail temperature ought to remain within the specified thermal limit (in theory), but this benefit comes at the cost of reduced sail accelerations and terminal speeds for higher values of $\varepsilon$.

There are two vital points to bear in mind concerning the above studies. The peak reflectance was governed by not only the grazing angle but also the chosen wavelengths. Given that real-world sails would be confronted with sail stabilization and control as well as the broadband SEDs of astrophysical sources, it appears unlikely that the fiducial values of $\Sigma_s$ and $\varepsilon$ will be attained in practice. It is instructive to gauge how $\beta_T$ will change if more conservative choices of the parameters are adopted. Upon substituting $\Sigma \approx 10^6\, \Sigma_0$ \citep{McAl18},\footnote{\url{https://www.radioactivity.eu.com/site/pages/Gamma_Attenuation.htm}} $\varepsilon \approx 1$ and $T_s \approx 400$ K in (\ref{NonrelTV}), we find that the new values of $\beta_T$ are reduced by three orders of magnitude compared to the prior estimates of this quantity (for a given $L_\star$).

In turn, we note that the acceleration distance and time derived in (\ref{SSecAccLS}) would also be duly modified. Let us work with the ansatz $v_F = \delta v_\infty$, implying that $\delta < 1$ constitutes the fraction of the terminal speed achieved. In this event, we find that the asymptotic values are $\Delta r \sim \delta^2 d_0$ and $\Delta t \sim 2 \delta d_0/v_\infty$. Thus, adopting the parametric choices outlined in the previous paragraph, after using (\ref{dzero}) and (\ref{vinf}), we find that $\Delta r$ increases by a factor of $\sim 5$ whereas $\Delta t$ is elevated by four orders of magnitude, provided that $\delta$ is held fixed. 

Hence, it seems necessary to view the estimates in Table \ref{TabLS} as highly optimistic, suggesting that at least some of them (e.g., GRBs and LXRBs) have to be downgraded by a few orders of magnitude. However, for certain high-energy sources - such as massive stars, AGNs \citep{VF09,BT17}, supernovae \citep{BW17}, and gamma-ray afterglows \citep{Pir05,KZ15} - a reasonably high fraction of photons are emitted at near-ultraviolet, optical and infrared wavelengths. For this class of systems, it is conceivable that the estimates in Table \ref{TabLS} are not vastly inaccurate. Moreover, we caution that our prior conclusions regarding $\beta_T$ were based on \emph{current} human technology. In principle, therefore, if a technologically advanced species is capable of absolute sail control and utilizes sophisticated nanomaterials, perhaps it might have the capacity to attain speeds that are not very far removed from those listed in Table \ref{TabLS} for some objects. 

\subsection{Constraints on the source environment}\label{SSecSourceC}
During the phase where the light sail is accelerated to its final velocity of $v_F$ in the vicinity of the astrophysical source, there are several key constraints imposed by the ambient gas and dust present in the environment. 

For starters, the following condition must hold true in order to prevent significant slow-down via the cumulative accrual of gas \citep{BL18}.
\begin{equation}\label{GasAcc}
    1.4 m_p \int_{d_0}^{r_F} n(r)\,dr < \Sigma_s,
\end{equation}
where $m_p$ is the proton mass, $n(r)$ represents the number density and $\Delta r$ is the acceleration distance estimated in (\ref{AccD}); the factor of $1.4$ accounts for the contribution of helium to the mass density of the gas. In this section, we will assume that the gas density obeys $n(r) \approx n_0 (d_0/r)^2$, which constitutes a reasonable assumption for certain astrophysical sources such as massive stars \citep{BD18}, thereby simplifying (\ref{GasAcc}) to
\begin{equation}
    1.4 m_p n_0 d_0 \left(\frac{v_F}{v_\infty}\right)^2 \lesssim \Sigma_s,
\end{equation}
after making use of (\ref{NRVelEOM}). Upon further simplification, the above equation reduces to
\begin{equation}\label{nzeromax}
    n_0 \lesssim 2.9 \times 10^8\,\mathrm{m}^{-3}\,\left(\frac{T_s}{300\,\mathrm{K}}\right)^4 \left(\frac{\varepsilon}{0.01}\right)^{-1} \left(\frac{\beta_F}{0.1}\right)^{-2}.
\end{equation}
In comparison, note that the characteristic value of the number density in the local interstellar medium (ISM) is around $10^6$ m$^{-3}$. There are two striking features that emerge from (\ref{nzeromax}): it does not depend on the luminosity of the source nor does it depend on the area density of the light sail. However, this statement is valid only if $\beta_F$ is held \emph{fixed}. Instead, if we presume that $\beta_F = \delta \beta_T$, we can utilize (\ref{NonrelTV}) to accordingly obtain
\begin{eqnarray}\label{nzeroterm}
&& n_0 \lesssim 3.3 \times 10^{12}\,\mathrm{m}^{-3}\,\delta^{-2} \left(\frac{L_\star}{L_\odot}\right)^{-1/2} \left(\frac{\Sigma_s}{\Sigma_0}\right) \nonumber \\
&& \hspace{0.5in} \times\,\left(\frac{T_s}{300\,\mathrm{K}}\right)^2 \left(\frac{\varepsilon}{0.01}\right)^{-1/2}.
\end{eqnarray}
Thus, it is evident that $n_0$ increases monotonically with $\Sigma_s$, whereas it declines when $L_\star$ is increased, both of which seem consistent with expectations.

Another major process responsible for the damage of light sails is ablation caused by impacts with dust grains. The limit on mass ablation is constructed from \citet[Equation 13]{BL18}, thereby yielding
\begin{equation}
  \frac{1.4 m_p \chi \varphi_{dg} \bar{m}}{\mathcal{U}}\int_{d_0}^{r_F} n(r) v^2(r)\,dr < \Sigma_s,   
\end{equation}
wherein $\chi \sim 0.2$ is the fraction of kinetic energy of the dust grain used to vaporize the sail material, $\varphi_{dg}$ is the dust-to-grain mass ratio, $\bar{m}$ is the mean atomic weight of the ablated material, and $\mathcal{U}$ is the vaporization energy. In formulating this expression, it was assumed that the dust grains are moving at much lower speeds than the light sail. After simplifying the integral, we end up with 
\begin{equation}
  \frac{0.7 m_p \chi \varphi_{dg} \bar{m} n_0 d_0 v_\infty^2}{\mathcal{U}} \left(\frac{v_F}{v_\infty}\right)^4 \lesssim \Sigma_s,   
\end{equation}
and we will tackle the case where $\beta_F = \delta \beta_T$. Using this scaling, the above equation is expressible as
\begin{eqnarray}\label{nzerotermv2}
&& n_0 \lesssim 1.4 \times 10^{12}\,\mathrm{m}^{-3}\,\delta^{-4} \left(\frac{L_\star}{L_\odot}\right)^{-1} \left(\frac{\Sigma_s}{\Sigma_0}\right)^2 \left(\frac{\chi}{0.2}\right)^{-1} \nonumber \\ 
&& \hspace{0.5in} \times\,\left(\frac{\varphi_{dg}}{0.01}\right)^{-1} \left(\frac{\bar{m}}{12\,m_p}\right)^{-1} \left(\frac{\mathcal{U}}{4\,\mathrm{eV}}\right).
\end{eqnarray}
A more comprehensive analysis of the drag as well as the ablation caused by dust grains and gas on weakly relativistic light sails has been undertaken in the context of the ISM by \citet{HLBL}.

The constraints on $n_0$ set by the astrophysical source environment are jointly embodied by (\ref{nzeroterm}) and (\ref{nzerotermv2}). If all the other parameters are held fixed, we note that (\ref{nzerotermv2}) constitutes the more stringent constraint for $L_\star > L_c$, whereas (\ref{nzeroterm}) becomes the more crucial constraint in the regime where $L_\star < L_c$. The critical luminosity $L_c$ that demarcates these two regimes is
\begin{eqnarray}
&& L_c \approx 0.17 L_\odot\,\delta^{-4} \left(\frac{\Sigma_s}{\Sigma_0}\right)^2 \left(\frac{T_s}{300\,\mathrm{K}}\right)^{-4} \left(\frac{\varepsilon}{0.01}\right) \left(\frac{\chi}{0.2}\right)^{-2} \nonumber \\
&& \hspace{0.5in} \times\,\left(\frac{\varphi_{dg}}{0.01}\right)^{-2} \left(\frac{\bar{m}}{12\,m_p}\right)^{-2} \left(\frac{\mathcal{U}}{4\,\mathrm{eV}}\right)^2.
\end{eqnarray}
Hence, if all the parameters are held fixed at their fiducial values, we find that $L_\star > L_c$ is valid for most astrophysical objects of interest provided that $\delta$ is not much smaller than unity. In other words, the primary constraint on $n_0$ is apparently set by (\ref{nzerotermv2}). We will, therefore, use this result in our subsequent analysis.

The constraint on the number density translates to a limit on the mass-loss rate ($\dot{M}_\star$) of the source via
\begin{equation}
  \dot{M}_\star \approx \Omega r^2 \rho_w(r) u_w(r),  
\end{equation}
under the assumption of spherical symmetry. Note that $\Omega$ denotes the solid angle over which the mass-loss rate occurs, whereas $\rho_w(r)$ and $u_w(r)$ are the mass density and the velocity of the wind. At distances $> d_0$, we will suppose that $u_w(r)$ remains approximately constant, which appears to be reasonably valid for stars \citep{VKL00,GV18}. We specify $r = d_0$ and utilize $\rho(d_0) = 1.4 m_p n_0$ in parallel with (\ref{nzeroterm}), thus arriving at
\begin{eqnarray}\label{MLRmax}
&& \dot{M}_\star \lesssim 2 \times 10^{-10}\,M_\odot\,\mathrm{yr}^{-1} \delta^{-4}\left(\frac{\Omega}{4\pi}\right) \left(\frac{u_w}{u_\odot}\right) \left(\frac{\Sigma_s}{\Sigma_0}\right)^2 \nonumber \\ 
&& \hspace{0.4in} \times\,\left(\frac{T_s}{300\,\mathrm{K}}\right)^{-4} \left(\frac{\varepsilon}{0.01}\right)\left(\frac{\chi}{0.2}\right)^{-1}\left(\frac{\varphi_{dg}}{0.01}\right)^{-1} \nonumber \\
&& \hspace{0.4in} \times\,\left(\frac{\bar{m}}{12\,m_p}\right)^{-1} \left(\frac{\mathcal{U}}{4\,\mathrm{eV}}\right),
\end{eqnarray}
In comparison, the current solar mass-loss rate is given by $\dot{M}_\odot \approx 2 \times 10^{-14} M_\odot\,\mathrm{yr}^{-1}$ \citep{Lin19}. Here, we have opted to normalize $u_w$ by $u_\odot = 500$ km/s, as it corresponds to the solar wind speed near the Earth \citep{Mar06}. The most striking aspect of (\ref{MLRmax}) is the fact that $L_\star$ is absent therein, which implies that the upper bound on $M_\star$ is independent of the source luminosity.

Next, we shall direct our attention to massive stars. Observations indicate that the terminal value of $u_w$ (denoted by $u_\infty$) is close to the escape speed ($v_\mathrm{esc}$) from the star \citep{VKL01,CS11}. Thus, it is possible to determine $u_\infty$, by utilizing the relationship $u_\infty \approx 1.3 v_\mathrm{esc}$ \citep{VKL00}, as follows:
\begin{equation}\label{vwrel}
  \frac{u_\infty}{u_\odot} \approx \left(\frac{M_\star}{M_\odot}\right)^{0.22},
\end{equation}
where we have employed the mass-radius relationship for massive stars \citep[pg. 320]{DK91}. By combining this relationship with (\ref{MLRmax}), we have obtained a heuristic upper bound on the stellar mass-loss rate that permits the efficient functioning of light sail acceleration. The empirical mass-loss rates for massive stars exhibit significant scatter and depend on a number of parameters such as the pulsation period, gas-to-dust mass ratio, and the luminosity \citep{GvL17}. We shall, however, adopt the simple prescription provided in \citet[Equation 3]{BD18} for massive stars at their end stages, which is expressible as
\begin{equation}
    \dot{M}_\star \approx 2.8 \times 10^{-25}\,M_\odot\,\mathrm{yr}^{-1}\, \left(\frac{L_\star}{L_\odot}\right)^{3.92}.
\end{equation}
In the case of intermediate mass stars, we adopt the mass-luminosity scaling from \citet[Table 3]{ESS15} and combine it with (\ref{MLRmax}) and (\ref{vwrel}) to arrive at
\begin{equation}
    M_\mathrm{max} \approx 9\,M_\odot\,\delta^{-0.38},
\end{equation}
with the rest of the parameters in (\ref{MLRmax}) held fixed at their characteristic values. The relevance of $M_\mathrm{max}$ stems from the fact that stars with $M_\star \gtrsim M_\mathrm{max}$ are potentially incapable of accelerating light sails to their terminal speeds without causing excessive damage in the process. The above expression suggests that smaller choices of $\delta$ can increase this threshold to some degree; for instance, if we choose $\delta \sim 0.1$, we end up with $M_\mathrm{max} \approx 21.6 M_\odot$.

There is another method by which we can deduce $M_\mathrm{max}$. By inspecting (\ref{vwrel}), we see that $u_\infty/u_\odot \approx 2$ for a star with mass $\sim 10 M_\odot$. By substituting this relation in (\ref{MLRmax}), we arrive at $\dot{M}_\star \lesssim 4 \times 10^{-10}\,M_\odot\,\mathrm{yr}^{-1} \delta^{-4}$. Hence, if we specify the range $\delta \approx 0.1$-$0.5$, we end up with $\dot{M}_\star \lesssim 6.4 \times 10^{-9}-4 \times 10^{-6}\,M_\odot\,\mathrm{yr}^{-1}$. Upon comparing these maximal mass-loss rates with those observed for O- and B-type stars \citep[Tables 1 and 2]{KCP18}, we determine that $M_\mathrm{max} \sim 10 M_\odot$. This estimate for the maximum stellar mass is consistent with the one obtained in the prior paragraph.

It is, however, necessary to appreciate that the ambient gas density and the mass-loss rate associated with massive stars (or AGNs) is not spherically symmetric because they exhibit a strong angular dependence relative to the rotation axis of the central object \citep{PVN08,NS14}. Hence, through the selection of launch sites where the density of gas and dust is comparatively lower, the above limit on $M_\mathrm{max}$ could be enhanced to a significant degree. We will not present an explicit estimate of this boost herein because of the inherent complexity of mass loss from massive stars.

The next astrophysical objects of interest that we delve into are SNe. The ejecta produced during the explosion move at typical speeds of $\sim 0.1 c$ \citep{KFM14,BW17}. By substituting this estimate for $u_w$ in (\ref{MLRmax}), we end up with $\dot{M}_\star \lesssim 1.2 \times 10^{-8}\,M_\odot\,\mathrm{yr}^{-1}\, \delta^{-4}$ when all the other parameters are held fixed. In comparison, the mass-loss rate of the progenitor just prior to the explosion is $\sim 0.01$-$0.1\,M_\odot\,\mathrm{yr}^{-1}$ \citep{KGA12} and it increases by a few orders of magnitude during the explosion. Hence, unless $\delta$ is sufficiently small, it is likely that SNe will cause significant damage to light sails situated in their vicinity. 

Lastly, we turn our attention to AGNs.\footnote{We will not tackle GRBs because they are transient events and do not therefore achieve speeds close to their asymptotic values within the time period these phenomena are functional.} Two contrasting phenomena are at work, namely, the inflow of gas that powers SMBHs and feedback-driven outflows \citep{VCB05,Fab12,DZ18}. These two processes are not mutually exclusive and are simultaneously at play in regions such as the AGN torus, thereby rendering modeling very difficult \citep{HA18}. Hence, for the sake of simplicity, we will suppose that the accretion occurs almost entirely within the Bondi radius ($r_B$), whose magnitude is given by \citep[Equation 1]{DAF03}:
\begin{equation}
    r_B \approx 4.6 \times 10^{-2}\,\mathrm{pc}\,\left(\frac{M_\mathrm{BH}}{10^6\,M_\odot}\right) \left(\frac{T_\mathrm{gas}}{10^7\,\mathrm{K}}\right)^{-1},
\end{equation}
where $T_\mathrm{gas}$ represents the temperature of the gas. This approach is consistent with the fact that AGN-driven outflows may play important roles at distances as small as $\sim 0.1$ pc \citep{ALB94,HTF16}. By comparing this result with (\ref{dzero}), after using (\ref{EddLum}) and assuming an Eddington factor of roughly unity \citep{MRG04}, we find $d_0 > r_B$ for SMBHs. Hence, we will restrict ourselves to the consideration of outflows.

The accretion of gas in AGNs is accompanied by wide-angle (i.e., non-collimated) outflows whose velocities vary widely. Although many quasars exhibit outflows with speeds of $\sim 0.1 c$ \citep{Kro99,MAB09,TMV15}, observations of other AGNs have identified winds and outflows at $\lesssim 0.01 c$ \citep[Section 2.3]{Fab12}. Upon substituting the optimistic case given by $u_w \sim 0.1 c$ into (\ref{MLRmax}), we end up with
\begin{equation}\label{MconsAGN}
\dot{M}_\star \lesssim 1.2 \times 10^{-8}\,M_\odot\,\mathrm{yr}^{-1}\, \delta^{-4}.
\end{equation}
In order to model the outflow mass-loss rate, we will employ a simple prescription, namely, that the outflow rate is proportional to the inflow (i.e., accretion) rate; the latter, in turn, is modeled using the Eddington accretion rate \citep{Shen13}. The proportionality constant $\zeta$ exhibits significant scatter - it ranges between $\lesssim 0.1$-$1000$ \citep{CK12}, although values of $\zeta \sim 100$ are seemingly more common \citep{KP09,DQM12,HTF16}. As per the preceding assumptions, the mass-loss rate arising from AGN outflows is expressible as
\begin{eqnarray}
&& \dot{M} \approx 2.2\,M_\odot\,\mathrm{yr}^{-1}\,\Gamma_e \left(1 - \epsilon_\mathrm{BH}\right) \left(\frac{M_\mathrm{BH}}{10^6\,M_\odot}\right) \nonumber \\ && \hspace{0.5in} \times\, \left(\frac{\zeta}{100}\right) \left(\frac{\epsilon_\mathrm{BH}}{0.1}\right)^{-1},
\end{eqnarray}
where $\Gamma_e$ is the Eddington ratio and $\epsilon_\mathrm{BH}$ represents the radiative efficiency of the SMBH. Hence, by comparing this expression with (\ref{MconsAGN}), we see that AGN outflows could cause significant damage to light sails in the event that $\delta$ is not much smaller than unity.

In view of the preceding discussion, it would appear as though there are noteworthy hindrances to deploying light sails in the vicinity of many high-energy astrophysical objects. However, there exist at least two avenues by which the aforementioned issues are surmountable in principle. First, by carefully selecting the timing at which the light sail is ``unfurled'', one might be able to operate in an environment where most of the ambient gas and dust has been cleared out (e.g., by shock waves), thus leaving behind radiation pressure to drive the spacecraft. Second, as we have seen, most of the hindrances arise from high ambient gas and dust densities. Hence, if the spacecraft is equipped with a suitable system to deflect these particles (provided that they possess a finite electrical charge or dipole moment) by means of electric or magnetic forces, one may utilize this device to prevent impacts and the ensuing ablation. 

This physical principle is essentially identical to the one underlying magnetic \citep{ZA91} and electric \citep{Jan04} sails, which are reliant upon the deflection of charged particles and the consequent transfer of momentum to the spacecraft. Thus, not only could one potentially bypass the dangers delineated thus far but also achieve a higher final speed in the process, albeit under ideal circumstances. We will not delve into this topic further as we briefly address electric sails in Sec. \ref{SecESail}. Likewise, it might also be feasible in principle to utilize an interstellar ramjet \citep{Bus60,Craw90,BG17} for the dual purposes of scooping up particles and gainfully employing them to attain higher speeds in the process.

We have not considered the slow-down arising from the hydrodynamic drag herein. This is because, as we shall demonstrate in Sec. \ref{SSecISM}, the drag force is potentially less effective in comparison to slow-down arising from the direct accumulation of gas; in particular, the reader is referred to (\ref{DistSD}) and (\ref{DgSD}) for the details. In a similar vein, we have not tackled the damage from sputtering as it contributes to the same degree as slow-down from gas accumulation \citep{BL18}; see also (\ref{DistSD}) and (\ref{DistSpu}) in the following section.

Finally, we turn our attention to the cascades caused by the impact of high-energy photons and particles with energies $\gtrsim 1$ GeV. Laboratory experiments have determined that each cascade displaces $\lesssim 10^3$ atoms when the colliding particle has energies of order of GeV \citep[pp. 77-130]{Was17}. We denote the average number flux of particles (taken here to be photons) in the source environment with $\gtrsim 1$ GeV energies by $\bar{F}_\mathrm{GeV}$. If a fraction $\mu_\mathrm{GeV}$ of particles trigger cascade formation, the ensuing constraint on the particle flux is expressible as
\begin{equation}
 \bar{F}_\mathrm{GeV} \cdot \mu_\mathrm{GeV} \cdot 10^3 \cdot \bar{m} \cdot \Delta t \lesssim \Sigma_s,
\end{equation}
where $\Delta t \sim 2 \delta d_0/v_\infty$ is the asymptotic acceleration time to reach the requisite final speed, after which the sail could be folded or discarded (see Sec. \ref{SSecISM}). After substituting the appropriate quantities, we arrive at
\begin{eqnarray}\label{PFluxCon}
&& \bar{F}_\mathrm{GeV} \lesssim 5.5 \times 10^{13}\,\mathrm{m}^{-2} \mathrm{s}^{-1}\,\delta^{-1} \mu_\mathrm{GeV}^{-1} \left(\frac{L_\star}{L_\odot}\right)^{-1/4} \left(\frac{\Sigma_s}{\Sigma_0}\right)^{1/2} \nonumber \\
&& \hspace{0.5in} \times\,\left(\frac{\varepsilon}{0.01}\right)^{-3/4} \left(\frac{T_s}{300\,\mathrm{K}}\right)^{3} \left(\frac{\bar{m}}{12\,m_p}\right)^{-1}.
\end{eqnarray}
Note that the photon flux obeys $F(r) = F_0 \left(d_0/r\right)^{-2}$, where $F_0$ signifies the flux at $r = d_0$. Therefore, the average flux during the acceleration phase is given by
\begin{equation}
\bar{F} \approx \frac{1}{\Delta r} \int_{d_0}^{r_F} F(r)\,dr \approx F_0 \left(1 - \delta^2\right),
\end{equation}
where the last equality follows after employing the definitions of $\Delta r$ and $r_F$. Even for a reasonably high value of $\delta = 0.3$, we see that $\bar{F} \approx F_0$, owing to which this relationship is adopted for all wavelength ranges. We denote the fraction of the total luminosity comprising photons with energies $\gtrsim 1$ GeV by $\kappa_\mathrm{GeV}$. Hence, $\tilde{L}_\star = 4\pi d_0^2 \bar{F}_\mathrm{GeV} (1\,\mathrm{GeV})/\kappa_\mathrm{GeV}$ represents a heuristic upper bound on the total luminosity of the source. Hence, by employing (\ref{PFluxCon}), we end up with
\begin{eqnarray}
&& \tilde{L}_\star \lesssim 0.19 L_\odot\,\delta^{-1} \mu_\mathrm{GeV}^{-1} \kappa_\mathrm{GeV}^{-1} \left(\frac{L_\star}{L_\odot}\right)^{3/4} \left(\frac{\Sigma_s}{\Sigma_0}\right)^{1/2} \nonumber \\
&& \hspace{0.4in} \times\,\left(\frac{\varepsilon}{0.01}\right)^{1/4} \left(\frac{T_s}{300\,\mathrm{K}}\right)^{-1} \left(\frac{\bar{m}}{12\,m_p}\right)^{-1}.
\end{eqnarray}
By dropping the tilde (i.e., setting $\tilde{L}_\star = L_\star$), we can invert the above relationship to estimate an upper bound on the luminosity of the source as follows:
\begin{eqnarray}\label{LminPF}
&& L_\star \lesssim 1.3 \times 10^{-3} L_\odot\,\delta^{-4} \left(\mu_\mathrm{GeV} \kappa_\mathrm{GeV}\right)^{-4} \left(\frac{\Sigma_s}{\Sigma_0}\right)^{2} \nonumber \\
&& \hspace{0.4in} \times\,\left(\frac{\varepsilon}{0.01}\right) \left(\frac{T_s}{300\,\mathrm{K}}\right)^{-4} \left(\frac{\bar{m}}{12\,m_p}\right)^{-4}.
\end{eqnarray}
Hence, for fiducial choices of the free parameters, we see that most high-energy astrophysical objects do not outwardly appear to meet this criterion, albeit in the highly extreme (and unrealistic) limit of $\mu_\mathrm{GeV} \rightarrow 1$ and $\kappa_\mathrm{GeV} \rightarrow 1$. If we increase $\Sigma_s$ or $\varepsilon$ in accordance with Sec. \ref{SSecSailM}, we see that the upper bound on $L_\star$ increases further. At first glimpse, this formula appears counterintuitive because it appears to rule out most stars, although the feasibility of stellar sailing is well documented \citep{McIn04,Vul12}. The answer lies in the fact that $\kappa_\mathrm{GeV}$ is many orders smaller than unity for all stars \citep{BV89}, thereby ensuring that (\ref{LminPF}) preserves consistency with expectations.  

\subsection{Effects of the interstellar medium}\label{SSecISM}
We assume henceforth that the light sail enters the interstellar medium (ISM) at the velocity $v_F$; depending on the interval over which the source remains active, $v_F$ may be close to the terminal velocity as explained earlier. Once the light sail enters the ISM, it will be subject to impacts by gas, dust and cosmic rays. This subject has been extensively studied by \citet{HLBL} and \citet{HL17}, but we will adopt the heuristic analysis by \citet{BL18} instead.

The first effect that merits consideration is the slow-down engendered by the accumulation of gas by the light sail. The mean number density of protons in the ISM along the spacecraft's trajectory is denoted by $\langle{n}\rangle$ and normalized in terms of $10^6\,\mathrm{m}^{-3}$ as noted previously. The maximum distance that is traversed by the spacecraft prior to experiencing significant slow-down ($D_a$) is
\begin{equation}\label{DistSD}
    D_a \approx 2.8\,\mathrm{pc}\,\left(\frac{\langle{n}\rangle}{10^6\,\mathrm{m}^{-3}}\right)^{-1} \left(\frac{\Sigma_s}{\Sigma_0}\right),
\end{equation}
The next effect that we address is collisions with dust grains, as they cause mass ablation upon impact. The corresponding maximal distance ($D_d$) is expressible as
\begin{eqnarray}\label{DistAbl}
&& D_d \approx 5 \times 10^{-5}\,\mathrm{pc}\,\left(\frac{\langle{n}\rangle}{10^6\,\mathrm{m}^{-3}}\right)^{-1} \left(\frac{\Sigma_s}{\Sigma_0}\right) \left(\frac{v_F}{0.1\,c}\right)^{-2} \nonumber \\
&& \hspace{0.4in} \times\,\left(\frac{\mathcal{U}}{4\,\mathrm{eV}}\right) \left(\frac{\chi}{0.2}\right)^{-1} \left(\frac{\varphi_{dg}}{0.01}\right)^{-1} \left(\frac{\bar{m}}{12\,m_p}\right)^{-1} .
\end{eqnarray}
An alternative expression for $D_d$ at weakly relativistic speeds (i.e., for $v_F > 0.1c$) is derivable from \citet[Equation 29]{Ho17} as follows:
\begin{equation}
 D_d \approx 54.8\,\mathrm{pc}\,\left(\frac{\langle{n}\rangle}{10^6\,\mathrm{m}^{-3}}\right)^{-1} \left(\frac{\mathcal{R}_\mathrm{min}}{1\,\mathrm{nm}}\right)^{1/2}, 
\end{equation}
wherein $\mathcal{R}_\mathrm{min}$ is the minimum size of interstellar dust grains. It must be noted, however, that the above equation was derived specifically for very thin light sails. 

As the light sail moves through the ISM, it will experience hydrodynamic drag due to the ambient gas. At low speeds, the drag force is linearly proportional to the kinetic energy of the sail \citep{Draine}, but this scaling breaks down at higher speeds. The maximum distance that can be covered by a weakly relativistic light sail before major slow-down due to hydrodynamic drag ($D_g$) is estimated from \citet[Equation 28]{Ho17}:
\begin{eqnarray}\label{DgSD}
&& D_g \approx 4.3 \times 10^{4}\,\mathrm{pc}\,\left(\frac{\langle{n}\rangle}{10^6\,\mathrm{m}^{-3}}\right)^{-1} \left(\frac{\Sigma_s}{\Sigma_0}\right) \nonumber \\
&& \hspace{0.4in} \times\,\left(\frac{v_F}{0.1\,c}\right)^{2.6}\left(\frac{\Delta \ell}{0.1\,\mathrm{\mu m}}\right)^{-1},  
\end{eqnarray}
where $\Delta \ell$ represents the thickness of the light sail. The last effect that we shall tackle entails sputtering due to gas, as it causes the ejection of particles from the light sail and thereby depletes its mass. The maximum travel distance before sputtering becomes a major hindrance ($D_s$) is expressible as \citep[Equation 17]{BL18}:
\begin{equation}\label{DistSpu}
D_s \approx 3\,\mathrm{pc}\,\left(\frac{\langle{n}\rangle}{10^6\,\mathrm{m}^{-3}}\right)^{-1} \left(\frac{\Sigma_s}{\Sigma_0}\right)\left(\frac{\mathcal{Y}}{0.1}\right)^{-1} \left(\frac{\bar{m}}{12\,m_p}\right)^{-1},
\end{equation}
where $\mathcal{Y}$ represents the total sputtering yield, with the associated normalization factor chosen in accordance with \citet[Figure 10]{TMS94}. Aside from sputtering, mechanical torques arising from collisions with ambient gas can result in spin-up and subsequent rotational disruption. At high speeds, however, this mechanism is apparently less efficient than sputtering in terms of causing damage unless the thickness of the light sail is $<0.01\,\mu$m \citep[Figure 5]{HL19}.

An inspection of (\ref{DistSD})-(\ref{DistSpu}) reveals that the upper bound on the distance is potentially $\lesssim 1$ pc for the parameter space described in the previous sections. Hence, at first glimpse, it would appear as though light sails moving at high speeds are not capable of travelling over interstellar distances. There is, however, a crucial factor that has been hitherto ignored. If the sail is ``folded'' in some fashion (e.g., retracted or deflated) or dispensed with altogether, the area density will be elevated by orders of magnitude. To see why this claim is valid, we shall consider the limiting case wherein the payload mass is roughly equal to the sail mass.\footnote{This constitutes the limiting case because one of the underlying assumptions in the paper was $M_{pl} \lesssim M_s$.} The size of the sail is denoted by $\mathcal{L}_s$, whereas the density and size of the payload are $\rho_{pl}$ and $\mathcal{L}_{pl}$, respectively. As the case delineated above amounts to choosing $\Sigma_s \mathcal{L}_s^2 \approx \rho_{pl} \mathcal{L}_{pl}^3$, reformulating this equation appropriately yields
\begin{equation}\label{AreaAmp}
\left(\frac{\mathcal{L}_s}{\mathcal{L}_{pl}}\right)^2 \approx 1.8 \times 10^6\, \left(\frac{\mathcal{L}_s}{1\,\mathrm{km}}\right)^{2/3} \left(\frac{\rho_{pl}}{\rho_0}\right)^{2/3} \left(\frac{\Sigma_s}{\Sigma_0}\right)^{-2/3},
\end{equation}
where $\rho_{pl}$ has been normalized in units of $\rho_0 \approx 4.5 \times 10^2$ kg/m$^3$, namely, the mean density of the International Space Station.\footnote{\url{https://www.nasa.gov/feature/facts-and-figures}} The significance of (\ref{AreaAmp}) is a consequence of the fact that this represents the amplification of the effective area density (stemming from the decrease in cross-sectional area) provided that the sail is completely folded. In other words, one would need to replace $\Sigma_s$ with $(\mathcal{L}_s/\mathcal{L}_{pl})^2\, \Sigma_s$ in (\ref{DistSD})-(\ref{DistSpu}). Hence, by closing the light sail, it ought to be feasible in principle for the spacecraft to travel distances on the order of kiloparsecs without being subject to major damage from the impediments arising from the ISM.

Lastly, even if the sail is folded, the collision of high-energy particles with the spacecraft will trigger the onset of cascades and thereby pose radiation hazards to electronics (and perhaps organics) on board the spacecraft \citep{Sem09}. It was estimated by \citet{HLL18} that spacecrafts traveling at $\sim 0.1 c$ would exhibit atomic depletion rates of $\sim 10^{17}$ m$^{-3}$ yr$^{-1}$ due to cascades arising from cosmic-ray impacts. However, it should be noted that the atomic densities of most solid materials are $\sim 10^{29}$ m$^{-3}$. Hence, this factor is unlikely to be important during the passage through the ISM. 

However, if the \emph{time-averaged} flux of GeV particles during the journey is $\sim 10^7$ times higher than the cosmic-ray flux near the Sun, this issue may prevent interstellar travel across kpc distances at speeds of $\sim 0.1 c$. Given that most of the high-energy particle flux ought to exist toward the beginning of the voyage, namely, in the vicinity of the astrophysical source, it is not clear as to whether such a high average flux would be prevalent; although this subject does necessitate further study, a detailed analysis is beyond the scope of the paper.

\subsection{Number of sails per source}
Hitherto, we have focused on analyzing the constraints on a single light sail. It should be noted, however, that a sufficiently advanced technological species could opt to place many light sails in the vicinity of the source and accelerate them to high speeds. By doing so, they can take advantage of the economies of scale \citep{Stig}, as the extra cost per additional light sail ought to be relatively low. We will delve into this possibility briefly, and highlight a few of the accompanying caveats.

Earlier, we have commented that the spacecrafts are launched from $r = d_0$. Hence, we consider a sphere of this radius and introduce the variable $\phi_s$ to denote the fraction of the sphere's surface that is covered by the light sails. Hence, the total number of spacecrafts per source ($\mathcal{N}_s$) is roughly estimated as
\begin{equation}
 \mathcal{N}_s \approx \phi_s \left(\frac{4\pi d_0^2}{\mathcal{L}_s^2}\right).   
\end{equation}
By making use of (\ref{dzero}), the above expression simplifies to
\begin{eqnarray}\label{Nsail}
&& \mathcal{N}_s \approx 2.5 \times 10^{3} \left(\frac{\phi_s}{\phi_0}\right)\left(\frac{\varepsilon}{0.01}\right) \left(\frac{T_s}{300\,\mathrm{K}}\right)^{-4} \nonumber \\
 && \hspace{0.6in} \times\, \left(\frac{L_\star}{L_\odot}\right) \left(\frac{\mathcal{L}_s}{1\,\mathrm{km}}\right)^{-2},
\end{eqnarray}
where we have chosen to normalize $\phi_s$ in units of $\phi_0 \approx 3 \times 10^{-13}$; the latter parameter embodies the fraction of light blocked by geostationary and geosynchronous satellites orbiting Earth \citep[Section 3.1]{SN18}. In principle, we could adopt much higher values of $\phi_s$, such as $\phi_s \approx 10^{-4}$ \citep{SN18}, which would elevate $\mathcal{N}_s$ by many orders of magnitude. For a supernova with $L_\star \approx 10^9\,L_\odot$, if we hold all of the other parameters fixed at their fiducial values, we find that $\mathcal{N}_s \approx 2.5 \times 10^{12}$ for kilometer-sized light sails. Hence, at least in principle, it is possible to ensure that the number of spacecrafts launched per high-energy source considerably exceeds the number of stars in the Milky Way by tuning the free parameters in (\ref{Nsail}) accordingly. 

There are, however, a couple of limitations to bear in mind. First, if $\mathcal{N}_s$ becomes exceedingly high, transporting such a large number of light sails to the astrophysical source from the parent system would pose difficulties; alternatively, one may attempt to construct them \emph{in situ} but this is contingent on the availability of raw materials. Second, we have deliberately adopted a conservative value of $\phi_s$ in (\ref{Nsail}), but even this miniscule fraction has a certain risk of the spacecrafts colliding with one another \citep{LG19}, and potentially triggering a collisional cascade known as the ``Kessler syndrome'' \citep{KCP}. While this risk is probably not a major concern for $\phi_s \approx \phi_0$ \citep{BW09,DH18}, it will become increasingly prominent at higher choices of $\phi_s$ \citep{SKC}, unless it can be overcome by an exceptional degree of spacecraft coordination and control. 

\section{Electric Sails}\label{SecESail}
Aside from light sails, several other propulsion systems do not require the spacecraft to carry fuel on board \citep{Long11}. Here, we will focus on just one of them, namely, electric sails. The basic concept underlying electric sails is that they rely upon electrostatic forces to deflect charged particles, and consequently transfer momentum to the spacecraft in this process. The major design principles underlying electric sails were delineated in \citet{Jan04}, following which many other studies have been undertaken in this field \citep{TJ09,JTP10,QM10,SRK13,BMQ19}. Another option is to implement the deflection of charged particles and concomitant momentum transfer using magnetic forces \citep{ZA91,Free15,Gros}, but we shall not tackle this method of propulsion herein. It is conceivable that the net effectiveness of electric and magnetic sails is comparable for certain parametric choices \citep{PH16}. 

\subsection{Basic properties of electric sails}\label{SSecESTermV}
In order to determine the acceleration produced by electric sails, one must calculate the force per unit length ($dF_s/dz$) and the mass per unit length ($dM_s/dz$). The former is difficult to estimate because it entails a complex implicit equation \citep{JTP10}. However, to carry out a simplified analysis, it suffices to make use of \citet[Equation 8]{JA07} and \citet[Equation 3]{TJ09}. The force per unit length for the electric sail is expressible as
\begin{equation}\label{FperL}
    \frac{dF_s}{dz} \approx  2\mathcal{K} m_p n \left(v - u_w\right)^2 r_D,
\end{equation}
where $\mathcal{K}$ is a dimensionless constant of order unity and $r_D$ is the Debye length, which is defined as
\begin{equation}\label{WirePar}
    r_D = \sqrt{\frac{\epsilon_0 k_B T_e}{n e^2}},
\end{equation}
wherein $\epsilon_0$ is the permittivity of free space and $T_e$ signifies the electron temperature. In reality, (\ref{FperL}) has been simplified because we neglected a term that is not far removed from unity, as it would otherwise make the analysis much more complicated (see \citealt{LL20} for additional details); the resulting expression for the acceleration is functionally identical to that of \citet{Jan04}. In addition, the factor of $v-u_w$ occurs in lieu of $v$, because prior studies were solely concerned with the regime where $v \ll u_w$ was valid. The mass per unit length for the sail material is given by
\begin{equation}\label{MeshMass}
    \frac{dM_s}{dz} = \pi \mathcal{R}_s^2 \rho_s,
\end{equation}
where $\mathcal{R}_s$ and $\rho_s$ denote the radius and density of the wire that comprises the electric sail. In order to maintain the sail at a constant potential, an electron gun is required, but we will suppose that its mass is smaller than (or comparable to) the sail mass; this assumption is reasonably valid at large distances from the source \citep{LL20}. The acceleration can be calculated by dividing (\ref{FperL}) with (\ref{MeshMass}). 

There are, however, some major issues that arise even when it comes to analyzing the spherically symmetric case. First, the density profile does not always obey the canonical $n \propto r^{-2}$ scaling; instead, it varies across jets, winds or outflows associated with different astrophysical sources. For example, the classic Blandford-Payne model for winds from magnetized accretion discs obeys $n \propto r^{-3/2}$ \citep{BP82}, whereas the outflows from Seyfert galaxies are characterized by $n \propto r^{-\alpha}$ with $\alpha \approx 1$-$1.5$ \citep{BJK06,Beh09}. Second, the scaling of the temperature is also not invariant: the Blandford-Payne model yields a power-law exponent of $-1$ while the solar wind exhibits an exponent of roughly $-0.5$ near the Earth \citep{LIM11}. Lastly, the velocity $u_w$ is not independent of $r$ in the regime of interest (namely, $r \gtrsim d_0$), although it eventually reaches an asymptotic value (denoted by $u_\infty \neq 0$) as per both observations and models \citep{Par58,VT98,Bes10}. 

\begin{figure*}
$$
\begin{array}{cc}
  \includegraphics[width=7.5cm]{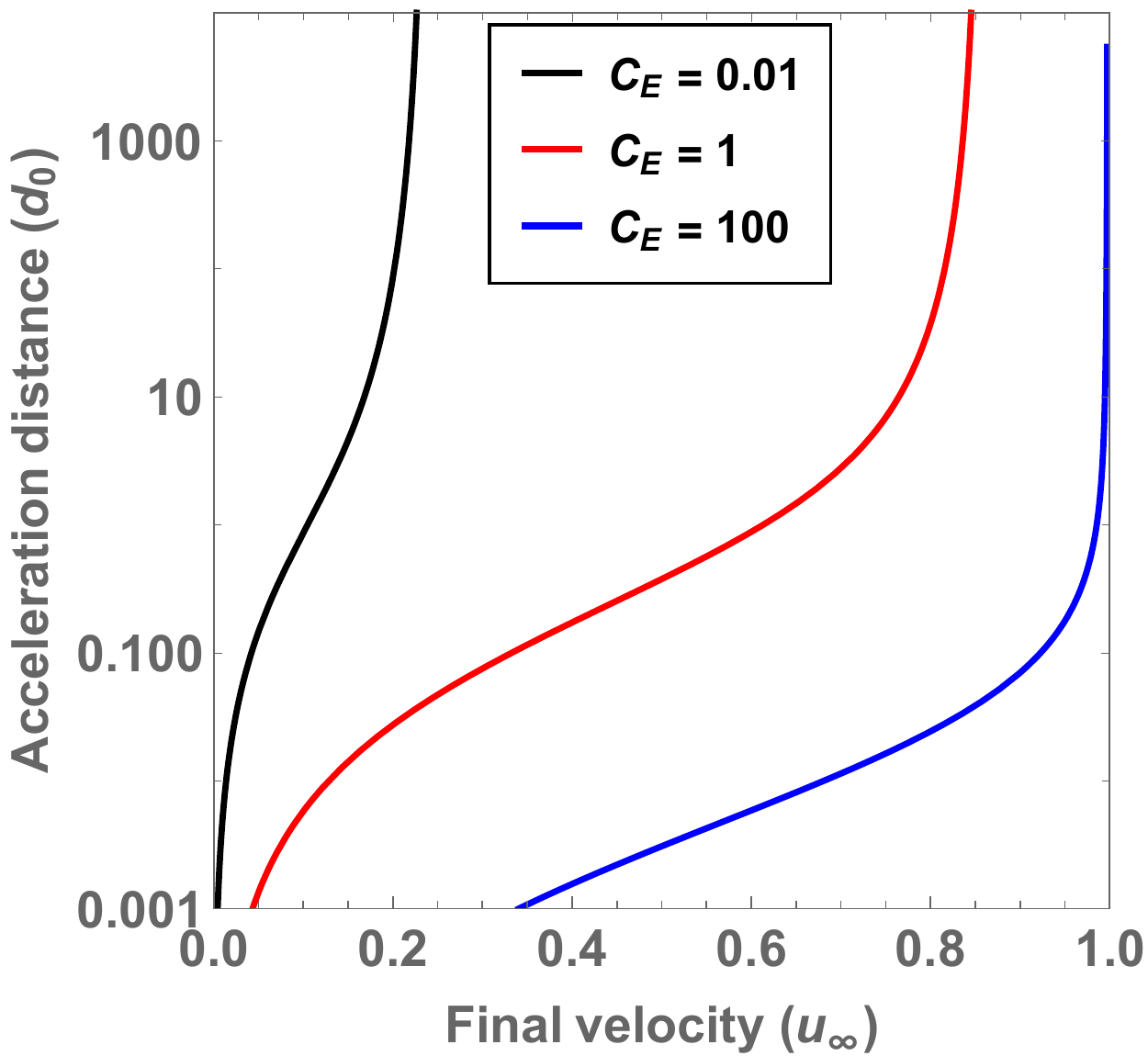} &  \includegraphics[width=7.5cm]{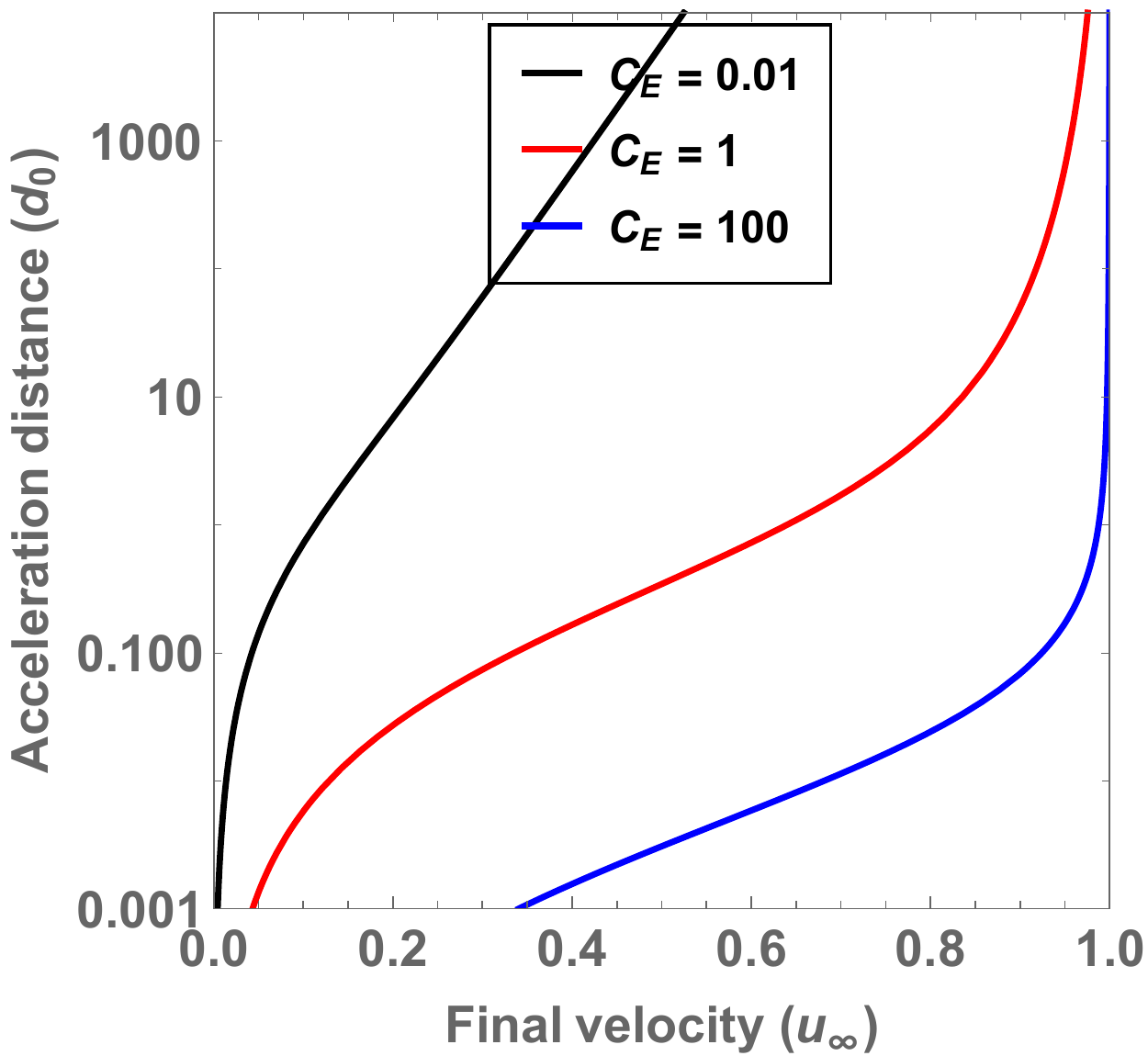}\\
\end{array}
$$
\caption{In both panels, the distance over which the electric sail must be accelerated (in units of the launch distance $d_0$) is shown as a function of the final velocity (in units of asymptotic wind speed $u_\infty$). The black, red and blue curves correspond to different choices of $\mathcal{C}_E$ in (\ref{vNRESail}). In the left-hand panel, we have chosen $\alpha = 2$ and $\xi = 0.5$, based on the parameters for stellar winds \citep{LL20}. We have specified $\alpha + \xi = 1.5$ in the right-hand panel, as this might be compatible with outflows detected in Seyfert galaxies.}
\label{FigAccDES}
\end{figure*}

Thus, this complexity stands in contrast to light sails, where the radiation flux falls off with distance as per the inverse square law. Hence, at first glimpse, it would appear very difficult to derive \emph{generic} properties of electric sails. We will, however, show that a couple of useful identities can nonetheless be derived. First, we consider the limiting case of $u_w \approx u_\infty$ as it constitutes a reasonable assumption at large enough values of $r$. We will also introduce the scalings $n \propto r^{-\alpha}$ and $T_e \propto r^{-\xi}$ and leave the exponents unfixed. To simplify our analysis, we employ the normalized variables $\tilde{v} \equiv v/u_\infty$, and $\tilde{r} \equiv r/d_0$. Using these relations along with (\ref{FperL})-(\ref{MeshMass}), we arrive at
\begin{equation}
  \tilde{a} \equiv \tilde{v} \frac{d \tilde{v}}{d \tilde{r}} \approx \mathcal{C}_E \left(\tilde{v} - 1\right)^2 \tilde{r}^{-(\alpha + \xi)/2},
\end{equation}
where $\mathcal{C}_E$ is a dimensionless constant that encapsulates the material properties of the electric sail as well as certain astrophysical parameters (e.g., source luminosity). In formulating the above expression, we have neglected the gravitational acceleration and hydrodynamic drag for reasons elucidated in Sec. \ref{SSecRelVTLS}. After integrating this equation, we end up with
\begin{equation}\label{vNRESail}
 \ln(1-\tilde{v}) + \frac{\tilde{v}}{1-\tilde{v}} \approx \frac{2\mathcal{C}_E}{\alpha + \xi -2} \left(1 - \tilde{r}^{-(\alpha + \xi - 2)/2}\right)  
\end{equation}
after specifying $\tilde{v} = 0$ at $\tilde{r} = 1$. 

Due to the uncertainty surrounding $\mathcal{C}_E$, $\alpha$ and $\xi$, we have plotted the normalized acceleration distance (given by $\tilde{r} - 1$) as a function of the final speed for various choices of these parameters in Fig. \ref{FigAccDES}. The right-hand panel, which satisfies the criterion $\alpha + \xi < 2$, yields results that are consistent with intuition. As we the sail speed approaches $u_\infty$, the acceleration distance diverges. On the other hand, the left-hand panel exhibits slightly unusual behavior that is dependent on $\mathcal{C}_E$. At large values of $\mathcal{C}_E$, we observe that the acceleration distance diverges in the limit of $\tilde{v} \rightarrow 1$ as before. However, when we have $\mathcal{C}_E \lesssim 1$, we noticed that the acceleration distance becomes singular at sail speeds that are conspicuously smaller than $u_\infty$. In other words, this implies that one cannot reach speeds close to $u_\infty$, irrespective of the distance travelled by the spacecraft. We will not estimate the acceleration time, because reducing (\ref{vNRESail}) to quadrature is not straightforward to accomplish.

Next, we shall formalize the above results by carrying out a mathematical analysis of (\ref{vNRESail}) for two distinct cases. The first scenario corresponds to $\alpha + \xi \leq 2$ and applying this limit to (\ref{vNRESail}) yields $\tilde{v} \rightarrow 1$. In other words, we end up with $v_\infty \approx u_\infty$ in this regime, which was also proposed in \citet{Jan04}. However, for a number of astrophysical systems (e.g., stellar winds) as well as classic theoretical models such as \citet{BP82}, we must address the case with $\alpha + \xi > 2$. By taking the limit of $\tilde{r} \rightarrow \infty$, the solution of (\ref{vNRESail}) is
\begin{equation}\label{vinfESail}
    \frac{v_\infty}{u_\infty} \approx 1 + \left[W\left(-\frac{1}{e} \exp(-\Upsilon)\right)\right]^{-1},
\end{equation}
where $W(x)$ is the Lambert $W$ function \citep{CGH96,VJC00} and we have introduced the auxiliary variable $\Upsilon = 2 \mathcal{C}_E/(\alpha + \xi - 2)$. Before analyzing (\ref{vinfESail}) in detail, it is important to recognize a subtle point.  By inspecting (\ref{vNRESail}), we see that $0 \leq \tilde{v} \leq 1$ because $\tilde{v} > 1$ would lead to the logarithmic function giving rise to non-real (i.e., complex) values. In other words, to ensure the existence of physically consistent solutions, we require $v_\infty/u_\infty \leq 1$ to be valid; in turn, this inequality states that the upper bound on $v_\infty$ is the terminal wind speed.

Depending on the magnitude of $\mathcal{C}_E$ (and therefore $\Upsilon$), there are two different regimes that require explication. First, let us consider the physically relevant scenario where $\mathcal{C}_E \gg 1$ holds true, which is potentially applicable to astrophysical sources with high luminosities. As this choice is essentially equivalent to taking the limit $\Upsilon \gg 1$, employing the latter yields
\begin{equation}\label{vTUpsInf}
    v_\infty \approx \left[\frac{\Upsilon + \ln\left(\Upsilon+1\right)}{\Upsilon + 1 + \ln\left(\Upsilon+1\right)}\right] u_\infty,
\end{equation}
which reduces further to $v_\infty \approx u_\infty$ when $\Upsilon \rightarrow \infty$. Next, suppose that we consider the opposite case wherein $\mathcal{C}_E \ll 1$. As this limit is tantamount to working with $\Upsilon \ll 1$, applying standard asymptotic techniques for the Lambert $W$ function near the branch point \citep{deB58,CGH96} leads to
\begin{equation}
    v_\infty \approx \left[\sqrt{2 \Upsilon} - \frac{4 \Upsilon}{3}\right] u_\infty,
\end{equation}
and substituting $\Upsilon \rightarrow 0$ implies that $v_\infty \rightarrow 0$. 

In summary, we found that choosing $\alpha + \xi \leq 2$ gave rise to $v_\infty = u_\infty$. On the other hand, for the physically pertinent case of $\alpha + \xi > 2$ and $\mathcal{C}_E \gg 1$, we approximately arrived at the same result; this is evident upon inspecting (\ref{vTUpsInf}). Hence, without much loss of generality, it is safe to assume that the terminal speed of electric sail for a given astrophysical object is set by the asymptotic value of the wind speed. In principle, one could also analyze the acceleration time and distance along the lines of Sec. \ref{SSecAccLS} and assess the constraints set by the source environment and the ISM.\footnote{As the electric sail is fundamentally composed of a wire mesh, it has a much smaller cross-sectional area than a light sail with the same dimensions, consequently facilitating the mitigation of damage caused by gas and dust.} However, we refrain from undertaking this study for two reasons: (i) many of the parameters as well as the scalings are non-universal and poorly determined, and (ii) the equation of motion is much more complicated, as seen from (\ref{vNRESail}), which makes subsequent analysis difficult.

\subsection{Terminal speeds of electric sails powered by astrophysical sources}\label{SSecESAstro}
For the aforementioned reasons, we shall confine ourselves to listing the observed values of $u_\infty$ for various astrophysical systems. It is natural to commence our discussion with stellar winds. By inspecting (\ref{vwrel}), it is apparent that $u_\infty$ only varies by a factor of $\sim 3$ even when $M_\star$ is increased by two orders of magnitude. Hence, insofar as stellar winds are concerned, the terminal wind speeds are on the order of $10^{-3}\, c$ in most cases; note that this statement also holds true for low-mass stars such as M-dwarfs \citep{DLMC,DHL17,DJL18,LL19}. Next, we consider SNe because the ejecta expelled during the explosion move at speeds of $\sim 0.1 c$, as noted in Sec. \ref{SSecSourceC}. Hence, this could serve as a rough measure of the final speeds that are attainable by electric sails in such environments. 

In the case of AGNs, there are two phenomena that need to be handled separately. The first are diffuse outflows that are characterized by $u_\infty \lesssim 0.1 c$ \citep[Equation 2.44]{Mer13}. These outflows have been identified in most quasars through the detection of broad absorption lines at ultraviolet wavelengths \citep{MCG95,GJB09,TCR13,KP15}. In contrast, relativistic jets from AGNs (i.e., blazars) typically exhibit Lorentz factors of $\mathcal{O}(10)$ \citep{PU92,Wor09,BMR19}; it is suspected that the observed jet emission is powered by magnetic reconnection \citep{SPG15}. Hence, at least in principle, it is possible for electric sails to attain such speeds provided that the relationship $v_\infty \approx u_\infty$ is still preserved.\footnote{A rigorous analysis of this complicated issue is beyond the scope of the paper, as it would entail the formulation and solution of the equations of motion for relativistic electric sails.} The Lorentz factors for jets arising from microquasars are usually of order unity \citep{MR99,RBMT}, suggesting that they also constitute promising sources for accelerating electric sails to relativistic speeds.

\begin{table}
\caption{Terminal momentum per unit mass achievable by electric sails near astrophysical objects}
\label{TabES}
\begin{tabular}{|c|c|}
\hline 
Source & Terminal momentum ($\gamma \beta$)\tabularnewline
\hline 
\hline 
Stars & $\sim 10^{-3}$\tabularnewline
\hline 
Supernovae & $\sim 0.1$\tabularnewline
\hline 
AGN outflows & $\sim 0.1$\tabularnewline
\hline 
Blazar jets & $\sim 10$\tabularnewline
\hline 
Microquasars & $\sim 1$\tabularnewline
\hline 
Pulsar wind nebulae & $\lesssim 10^4-10^5$\tabularnewline
\hline 
\end{tabular}
\medskip

{\bf Notes:} $\gamma \beta$ denotes the terminal momentum per unit mass. It is important to recognize that this table yields the \emph{maximum} terminal speeds attainable by electric sails, because it assumes that the terminal sail speeds approach the asymptotic values of the winds, outflows and jets. However, this assumption may not always be valid, as explained in Sec. \ref{SSecESTermV}. Lastly, we note that these values are fiducial, and a more complete analysis is provided in Sec. \ref{SSecESAstro}.
\end{table}

Pulsar wind nebulae (PWNe) will be the last example that we shall study here. PWNe comprise highly energetic winds that are powered by a rapidly rotating and highly magnetized neutron star \citep{GS06,KCL15}. The energy loss is caused by the magnetized wind emanating from the neutron star, and is expressible as \citep[Equation 2]{Sla17}:
\begin{equation}
    \dot{E} = - \frac{B_p R_p^6 \omega_p^4}{6 c^3} \sin^2\Theta,
\end{equation}
where $B_p$ is the dipole magnetic field strength at the poles, $R_p$ and $\omega_p$ are the radius and rotation rate of the pulsar, and $\Theta$ is the angle between the pulsar magnetic field and rotation axis. The minimum particle current ($\dot{N}$) that is necessary for the sustenance of a charge-filled magnetosphere is estimated using \citet[Equation 10]{GS06}, which equals
\begin{equation}
    \dot{N} = \frac{B_p R_p^3 \omega_p^2}{\mathcal{Z} e c},
\end{equation}
where $\mathcal{Z} e$ represents the ion charge; this relationship was first determined by \citet{GJ69}. The maximum Lorentz factor ($\gamma_\mathrm{max}$) that is achievable in pulsar winds occurs near the termination shock, the location at which the ram pressure of the wind and the ambient pressure in the PWN balance each other, and has been estimated to be \citep[pg. 2164]{Sla17}:
\begin{equation}
  \gamma_\mathrm{max} \approx 8.3 \times 10^6\,\left(\frac{\dot{E}}{10^{31}\,\mathrm{J}}\right)^{3/4} \left(\frac{\dot{N}}{10^{40}\,\mathrm{s}^{-1}}\right)^{-1/2}. 
\end{equation}
It is important to note, however, that $\gamma$ is typically on the order of $100$ just outside the light cylinder, which is defined as $c/\omega_p$ \citep[Section 4.4]{GS06}. The analysis of data from young PWNe in conjunction with spectral evolution models yielded bulk Lorentz factors of $\gamma \sim 10^4-10^5$ for the pulsar winds \citep[Table 2]{TT11}. It is worth noting that the characteristic synchrotron emission lifetime of particles in PWNe is $\sim 10^3$ yr \citep[Equation 10]{Sla17}. Most PWNe that have been detected are young (with ages of $\sim 10^3$ yr), but some PWNe discovered by the \emph{Suzaku} X-ray satellite have ages of $\sim 10^5$ yr and are apparently still fairly active \citep{BAD10}. Hence, the lifetime over which PWNe are functional may suffice to accelerate putative electric sails close to the bulk speeds of pulsar winds.

Lastly, another chief advantage associated with electric sails merits highlighting. Hitherto, we have seen that a variety of sources are capable of accelerating light sails or electric sails to relativistic speeds on the order of $0.1c$. However, after the spacecraft has been launched toward the target planetary system, it will need to eventually slow down and attain speeds of order of tens of km/s to take part in interplanetary maneuvers. Electric sails are a natural candidate for enforcing comparatively rapid slow down through the process of momentum transfer with charged particles in the ISM. More specifically, \citet{PH16} concluded that spacecrafts with total masses of $\sim 10^4$ kg could be slowed down from $0.05\,c$ to interplanetary speeds over decadal timescales by utilizing an electric sail.\footnote{In principle, stellar radiation pressure is also suitable for slowing down light sails near low-mass stars, as explicated by \citet{Forw84} and \citet{HH17}.}

\section{Conclusions}\label{SecConc}
In this work, we investigated the possibility of harnessing high-energy astrophysical phenomena to drive spacecrafts to relativistic speeds. In order to bypass the constraints imposed by the rocket equation, we focused on light sails and electric sails because: (i) neither of them are required to carry fuel on board, (ii) they possess the capacity to attain high speeds, and (iii) they are both relatively well-studied from a theoretical standpoint and successful prototypes have been constructed.

Our salient results are summarized in Tables \ref{TabLS} and \ref{TabES}. From these tables, it is apparent that speeds on the order of $\gtrsim 0.1 c$ may be realizable by a number of astrophysical sources, and Lorentz factors much greater than unity might also be feasible in certain environments. In the event that such ultrarelativistic speeds are realizable in actuality, it could be possible to undertake intergalactic exploration \citep{AS13,Ols15,Ols17,San18}. For example, if a Lorentz factor of $\sim 10^3$ is attained, and the spacecraft mostly travels at uniform velocity, it would be possible to reach the Andromeda galaxy (M31) in a span of $\sim 2.5 \times 10^3$ yrs as measured by an observer on board the spacecraft. 

We reiterate that the values presented in the aforementioned tables should be regarded as optimistic upper bounds. In the case of light sails, we carried out a comprehensive analysis of whether the astrophysical sources last long enough to permit the attainment of relativistic speeds as well as the constraints on sail materials, the source environment and the passage through the ISM. We concluded that all of these effects pose significant challenges, but could be overcome in principle through careful design. We also estimated the number of light sails that can be accelerated per source, and we determined that it may exceed the number of stars in the Milky Way under optimal conditions. When it came to electric sails, several additional uncertainties were involved, owing to which we restricted ourselves to estimating their maximum terminal speeds; these speeds are not necessarily achievable under realistic circumstances.

Our analysis entailed the following major caveats. First, we carried out the calculations in simplified (i.e., one-dimensional) geometries wherever possible, which constitutes an idealization for most time-varying astrophysical sources. Second, our analysis did not take numerous engineering constraints into account, with the exception of a temperate spacecraft temperature. In this context, there are many key issues such as maintaining sail stability and control, possessing requisite structural integrity, mitigating spacecraft charging,\footnote{Methods for alleviating charge accumulation and the torques arising from induced asymmetric charge distributions have been delineated in \citet{GW12,HL17}.} and ensuring broadband reflectance and minimizing absorptance (due to the nature of astrophysical SEDs) that are not tackled herein. In the same vein, we do not address economic costs and benefits of space exploration \citep{And04,Hos07,Kru10,Cap16} or the ethical and sustainability issues surrounding it \citep{Wil03,Pop08,HB09,SM16}, both of which are indubitably of the highest importance. In light of these facts, our work should therefore be viewed merely as a preliminary conceptual study of the \emph{maximum} terminal speeds that may be achievable by light or electric sails in the vicinity of high-energy astrophysical objects.

Aside from the obvious implications for humanity's own long-term future, our results might also offer some pointers in the burgeoning search for technosignatures. In particular, searches for technosignatures could focus on high-energy astrophysical sources, as they represent promising potential sites for technological species to position their spacecrafts; this complements the earlier notion that these high-energy phenomena constitute excellent Schelling points (see \citealt{Wri18} for a review). We caution, however, that the putative spacecrafts under consideration have a low likelihood of being detectable, due to the intrinsic temporal variability of high-energy astrophysical sources \citep{MSL11}. The best option may entail searching for: (i) radio signals in the vicinity of these sources, if the spacecrafts are communicating with one another, and (ii) megastructures \citep{Kar85,WCZ} such as Stapledon-Dyson spheres \citep{Stap37,Dys60} and ring worlds \citep{Niv70,McIn} in the vicinity of these objects \citep{Osm16,IDS18}.

Another option is to search for techosignatures of relativistic spacecraft as they traverse the ISM \citep{VHP77}. Some possibilities include the detection of cyclotron radiation emitted by magnetic sails \citep{Zub95}, extreme Doppler shifts caused by reflection from relativistic light sails \citep{GC13}, and radiation signatures generated by scattering of cosmic microwave background photons from the relativistic spacecraft \citep{YW18}. Even non-relativistic spacecrafts, provided that they are either sufficiently large or tightly clustered and numerous, could give rise to detectable infrared excesses \citep{Teo14,Osm19} in the manner of Stapledon-Dyson spheres.

Finally, it has been suggested since the 1960s that searches for probes and artifacts in our Solar system may represent a viable line of enquiry \citep{Bra60,Sag63}. A number of targets have been proposed in this context such as the neighborhood of the Solar gravitational lens \citep{Gil14}, Oort cloud and Kuiper belt objects \citep{Ger16}, the asteroid belt \citep{Pap78,Pap83}, surfaces of the Moon and Mars \citep{BW00,HK12,DW13,MaLi}, Earth-Moon Lagrange points \citep{FV80,VF83}, co-orbital near-Earth objects \citep{Ben19}, the upper atmosphere of the Earth \citep{Teo01,SL20}, and Earth's surface and subsurface environments \citep{Ark96,Dav12,JW18,SF19}. What remains unknown, however, is the probability of success for any of the aforementioned strategies, because it ultimately comes down to the question of how many technological species are extant in the Milky Way \citep{SS66,VD15,Cir18,ML19}.

\acknowledgments
We thank the reviewer for the detailed and insightful report that greatly aided in the improvement of the paper. This work was supported in part by the Breakthrough Prize Foundation, Harvard University's Faculty of Arts and Sciences, and the Institute for Theory and Computation (ITC) at Harvard University.


\end{document}